\documentclass[fleqn,usenatbib]{mnras}

\usepackage[T1]{fontenc}

\DeclareRobustCommand{\VAN}[3]{#2}
\let\VANthebibliography\thebibliography
\def\thebibliography{\DeclareRobustCommand{\VAN}[3]{##3}\VANthebibliography}

\usepackage{amsmath}
\usepackage{natbib}

\usepackage{soul}
\usepackage{color}
\usepackage{textcomp}

\usepackage{graphicx}

\usepackage{hyperref}

\newcommand{\chandra}{\emph{Chandra}}

\newcommand{\fu}{4U~1630$-$47}

\newcommand{\wsim}{\ensuremath{\sim}}

\newcommand{\degree}{\ensuremath{^\circ}}
\newcommand{\amin}{\ensuremath{'}}

\newcommand{\FU}{4U~1630$-$47}
\newcommand{\swift}{\emph{Swift}}
\newcommand{\apex}{\emph{APEX}}
\newcommand{\maxi}{\emph{MAXI}}
\newcommand{\HI}{H\,{\sc i}}

\title[DSH distance 4U~1630$-$47]{Dust scattering halo of 4U~1630$-$47: High resolution X-ray and mm observations constrain source and molecular cloud distances}

\author[E. Kalemci et al.]{E. Kalemci$^{1}$\thanks{E-mail: ekalemci@sabanciuniv.edu},
M. D\'iaz Trigo$^{2}$, E. Oztaban$^{1}$, A. A. Abbasi$^{1}$, T. Stanke$^{3}$,
\newauthor J. A. Tomsick$^{4}$, T. J. Maccarone$^{5}$, 
A. Sara\c{c}yakupo\u{g}lu$^{1}$, E. von Nussbaum$^{6,7}$,
\newauthor J. C. A. Miller Jones$^{8}$, B. Bah\c{c}eci$^{1}$
\\
$^{1}$Faculty of Engineering and Natural Sciences, Sabanc\i\ University, Orhanl\i-Tuzla, 34956, Istanbul, Turkey\\
$^{2}$ESO, Karl-Schwarzschild-Stra\ss{}e 2, D-85748, Garching bei M\"unchen, Germany\\
$^{3}$Max-Planck-Institut f\"ur Extraterrestrische Physik, Gie\ss{}enbachstra\ss{}e 1, Garching bei M\"unchen, DE\\
$^{4}$Space Sciences Laboratory, 7 Gauss Way, University of California, Berkeley, CA, 94720-7450, USA\\
$^{5}$Department of Physics \& Astronomy, Texas Tech University, Box 41051, Lubbock TX, 79409-1051 USA\\
$^{6}$Otto von Taube Gymnasium, Germeringer Stra\ss{}e 31, D-82131 Gauting, DE\\
$^{7}$Technische Universit\"at M\"unchen, D-80333 M\"unchen, DE\\
$^{8}$International Centre for Radio Astronomy Research -- Curtin University, GPO Box U1987, Perth, WA 6845, Australia
}

\date{Accepted XXX. Received YYY; in original form ZZZ}

\pubyear{2025}

\begin{document}
\label{firstpage}
\pagerange{\pageref{firstpage}--\pageref{lastpage}}
\maketitle

\begin{abstract}
 
We re-investigated the distance to the black hole X-ray binary \fu\ by analyzing its dust scattering halo (DSH) using high-resolution X-ray (\chandra) and millimeter (\apex) observations. Dust scattering halos form when X-rays from a compact source are scattered by interstellar dust, creating diffuse ring-like structures that can provide clues about the source's distance. Our previous work suggested two possible distances: 4.9 kpc and 11.5 kpc, but uncertainties remained due to low-resolution CO maps. We developed a new methodology to refine these estimates, starting with a machine learning approach to determine 3D representation of molecular clouds from the \apex\ dataset. The 3D maps are combined with X-ray flux measurements to generate synthetic DSH images. By comparing synthetic images with the observed \chandra\ data through radial and azimuthal profile fitting, we not only measure the source distance but also distinguish whether the molecular clouds are at their near- or far-distances. The current analysis again supported a distance of 11.5 kpc over alternative estimates. While the method produced a lower reduced $\chi^{2}$ for both the azimuthal and radial fits for a distance of 13.6 kpc, we ruled it out as it would have produced a bright ring beyond the \apex\ field of view, which is not seen in the \chandra\ image. The 4.85 kpc estimate was also excluded due to poor fit quality and cloud distance conflicts. The systematic error of 1 kpc, which arises due to errors in determining molecular cloud distances, dominates the total error. 

\end{abstract}


\section{Introduction}\label{sec:intro}

X-ray transients in outburst may span orders of magnitude in luminosity range and may temporarily become the brightest X-ray objects in the sky \citep{Kalemci22}. There is clear evidence for super-Eddington accretion in some ultraluminous X-ray sources \citep{Bachetti2014, Brightman2019}, with some debate over whether this is due to bona fide super-Eddington accretion or beaming. There is also tentative evidence that some Galactic black hole X-ray binaries reach super-Eddington phases \citep{Jin2024, Muno99, Uttley15, Revnivtsev2002}. However, uncertainties in distance, mass, and inclination angle prevent firm conclusions, with only the spectral properties being fully established in systems at high luminosities in a way that is consistent with radiation pressure being important.
Accurate distance measurements are not only the key to establishing reliable Eddington luminosity fractions, but they are also essential in determining the size scales, inclination angles and speeds of jets observed in black hole systems \citep{Burridge2025}.

Most of the X-ray transients are in the Galactic Plane behind copious amounts of dust and gas \citep[see][for both distance estimates and Galactic distribution]{CorralSantana16}. The extinction due to high column density makes it difficult to determine the distances using methods relying on optical telescopes. There are indirect methods to measure distances such as using state transition luminosities in X-rays \citep{Maccarone03_a, Vahdat2019}, or other X-ray spectral analysis methods \citep{Abdulghani24}. Radio observations may be used as well, including jet-parallax \citep{MillerJ09, Atri2020}, which is model-independent, and \HI\ absorption lines \citep{Chauhan21, Burridge2025}. \cite{Arnason2021} discusses how the methods fare when compared to the optical parallax measurements with the \emph{Gaia} mission. For bright sources behind moderate to high dust concentration, one can also analyze the radial profile of the halo caused by the scattering of X-rays from dust and determine the source distance in this manner \citep{Trumper73, Predehl00, Thompson09, Xiang11, Heinz15}.

In \citealt{Kalemci18} (hereafter K18), we utilized \chandra, \swift, and low-resolution CO maps to constrain the distance to the black hole transient \FU\ by fitting the dust scattering halo (DSH) radial profile. We reported two possibilities: 11.5 kpc and 4.9 kpc, favoring 11.5 kpc with the main scattering molecular cloud MC-79 being at the far distance estimate obtained using a Galaxy rotation curve model. In the same study, we asserted that the resolution of CO maps is the source of the main uncertainty in the method. Recently, to overcome the difficulties caused by low-resolution images, we obtained high-resolution mm maps of the region with the Atacama Pathfinder Experiment \citep[\apex,][]{Gusten2006} and developed an imaging-based method to determine both the source distance and the order of molecular clouds in the line of sight. In this article, we describe the methodology and present its first results applied to \FU.

\subsection{4U~1630$-$47}\label{subsec:fu}

After its discovery in 1976 (\cite{jones76}), \fu\ has been classified as a low-mass X-ray binary (LMXB), with recurrent quasi-periodic outbursts every 2-3 years. It was first classified as a black hole from its spectral and timing properties by \cite{Barret96}. It is a peculiar Galactic black hole transient (GBHT), often not following the typical hysteresis pattern in the hardness-intensity diagram \citep{Tomsick05, Abe05}. 

The source is located in the Galactic Plane with a high and varying absorption column density within and between spectral states \citep{Augusteijn01, Tomsick14}, making accurate estimates of its distance difficult. Through $IXPE$ observations, \fu\ has been revealed to have high polarization in X-rays \citep{Krawczynski23}. The distance to the source directly affects the interpretation of these measurements, as an accurate estimate of the distance can improve the reliability of the models describing the accretion dynamics and the observed polarization of the source.

\fu\ has also shown a very low luminosity ($L/L_{Edd}$=0.0003) soft state \citep{Tomsick14}. Based on the iron line features simultaneously observed with radio emission, a baryonic content for the jet has been proposed \citep{DiazTrigo13}, but other explanations remain \citep{Wang16, Neilsen14}. In K18, we tested the idea that some of the peculiarities of the source can be explained by local dust scattering through analysis of the dust scattering halos of the source observed with \chandra\ and \swift.

\subsection{Dust scattering halos}
\label{subsec:DSHintro}

Dust scattering halos \citep{Overbeck65, Mathis91, Predehl95} are diffuse rings of light that form when X-rays are scattered by interstellar dust grains along the line of sight to a distant source. In addition to the determination of the source distance mentioned above, studies of these halos have provided valuable information on spectral variations during eclipses \citep{Audley06}, the physical characteristics of dust grains and their distribution along the line of sight \citep{Corrales15, Xiang11}, and the measurement of X-ray extinction \citep{Predehl95, Corrales16}. For a recent review, see \citealt{Constantini2022}.

\begin{figure}
\includegraphics[width=\columnwidth]{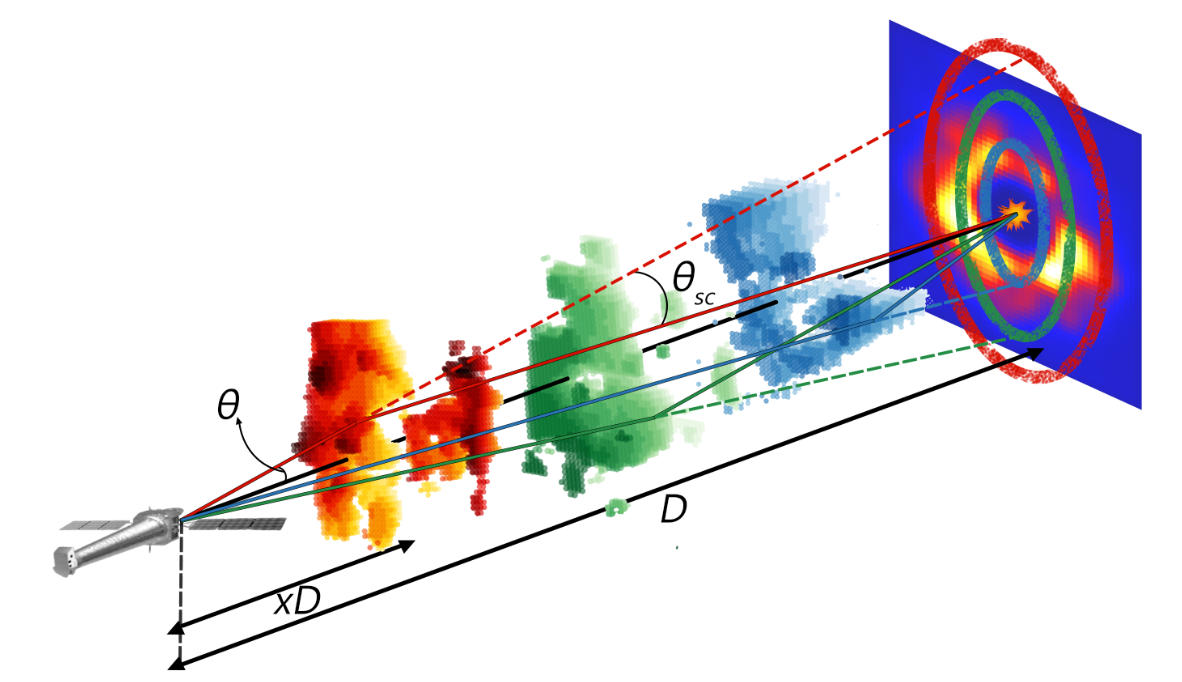}
\caption{\label{fig:dshgeom}
Figure depicting the formation of dust scattering halos using the 3D shapes of the molecular clouds. The source distance is given by $D$, and the distance to the dust cloud from the observer is given by $xD$. $\theta$ is the observed scattering angle, $\theta_{sc}$ is the physical scattering angle. The distances of 3D clouds in this figure are for better representation and do not coincide with actual distances in this work.
 }
\vspace{0.5 pt}
\end{figure}

When a source experiences outbursts (or flares) followed by an extended period of quiescence, the dust-scattered emission appears as distinct ring-like structures. This occurs because dust along the line of sight is concentrated in molecular clouds, and each ring structure corresponds to the delayed scattered emission from a single region of high dust density (see Figure~\ref{fig:dshgeom}). The delay is a result of the scattered X-rays taking a longer path compared to the direct X-rays observed by the telescope \citep[][and references therein]{Heinz16}. The DSH appears as uniform rings only if the scattering is also caused by uniform-density dust in the clouds. As explored in this work, the nonuniformity of clouds will be reflected in the azimuthal distribution of emission and absorption in the rings. 

\begin{table}
\centering
\caption{\swift\ Observations used in the analysis} 
\begin{tabular}{cccc}
\hline \hline

Date (MJD) & obsid & $N_{H}$ & $T_{in}$ \\
\hline \hline

57630.5 & 00031224020 & 13.99$\pm$0.21* & 1.295$\pm$0.025 \\
57635.3 & 00031224022 & 14.72$\pm$0.74 & 1.554$\pm$0.106 \\
57636.0 & 00031224023 & 13.77$\pm$0.14 & 1.481$\pm$0.021 \\
57637.1 & 00031224024 & 14.29$\pm$0.32 & 1.495$\pm$0.048 \\
57638.3 & 00031224025 & 13.14$\pm$0.19 & 1.582$\pm$0.033 \\
57639.3 & 00031224026 & 11.10$\pm$0.66 & 1.359$\pm$0.111 \\
57640.0 & 00031224027 & 9.060$\pm$1.10 & 1.284$\pm$0.320 \\
57641.5 & 00031224028 & 12.77$\pm$1.05 & 1.355$\pm$0.141 \\
57642.5 & 00031224029 & 13.37$\pm$0.14 & 1.565$\pm$0.023 \\
57643.4 & 00031224030 & 14.07$\pm$0.16 & 1.620$\pm$0.026 \\
57644.3 & 00031224031 & 14.53$\pm$0.24 & 1.583$\pm$0.036 \\
57645.9 & 00031224032 & 14.35$\pm$0.15 & 1.677$\pm$0.027 \\
57648.2 & 00031224033 & 10.83$\pm$0.48 & 1.349$\pm$0.079 \\
57654.1 & 00031224035 & 14.87$\pm$0.25 & 1.541$\pm$0.037 \\
57657.9 & 00031224036 & 13.85$\pm$0.24 & 1.470$\pm$0.038 \\
57658.5 & 00031224037 & 14.41$\pm$0.41 & 1.662$\pm$0.061 \\
57660.9 & 00031224038 & 13.30$\pm$0.18 & 1.519$\pm$0.028 \\
57663.6 & 00031224039 & 13.58$\pm$0.16 & 1.503$\pm$0.024 \\
57667.1 & 00031224040 & 14.03$\pm$0.17 & 1.650$\pm$0.029 \\
57669.5 & 00031224041 & 13.04$\pm$0.16 & 1.577$\pm$0.027 \\
57672.4 & 00031224042 & 12.40$\pm$0.41 & 1.672$\pm$0.080 \\
57675.6 & 00031224043 & 13.12$\pm$0.85 & 1.452$\pm$0.089 \\
57678.3 & 00031224044 & 14.10$\pm$0.14 & 1.741$\pm$0.027 \\
57682.5 & 00031224045 & 14.37$\pm$0.18 & 1.765$\pm$0.033 \\ \hline
\end{tabular}
\label{table:obs}
\small
*Errors are at 1$\sigma$
\end{table}

\section{Observations and Analysis}\label{sec:obs}

In this work, we utilized data from several observatories: in X-rays \chandra, \swift, and \maxi\ and in mm band \apex. We also used publicly available CO maps from the Bronfman survey \citep{Bronfman89} for a larger coverage of the molecular cloud distribution and the Southern Galactic Plane Survey \citep[SGPS,][]{McClure2005} to estimate the column density of neutral atomic hydrogen.

\subsection{X-ray imaging analysis}\label{subsec:chandra}

\begin{figure}
\includegraphics[width=\columnwidth]{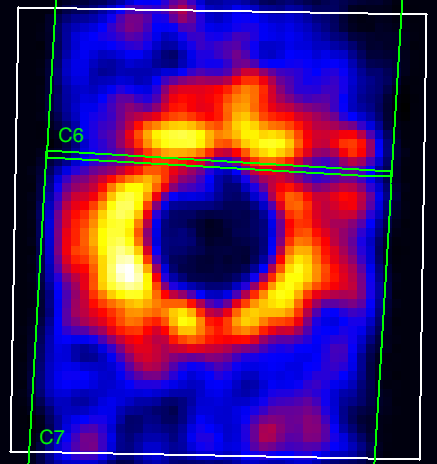}
\caption{\label{fig:chrebine2}
The rebinned and background-subtracted \chandra\ image in the 2.25--3.15 keV band. All point sources (including \fu) have been removed. A Gaussian smoothing with 3 pixel radius was applied. The white box shows the \apex\ field of view, and green boxes show \chandra\ ACIS-S chip boundaries with chip numbers shown at the lower left corners.
}
\vspace{0.5 pt}
\end{figure}

We used the same data set in K18 from \chandra\ (obsid 19004) to recreate the main dust scattering halo image. Fig.~\ref{fig:chrebine2} shows the background subtracted, point sources removed, and rebinned \chandra\ image (matching the \apex\ resolution) in the 2.25--3.15 keV band. The same figure also shows the \chandra\ ACIS-S chip boundaries as well as the \apex\ field of view. A larger image in three energy bands can be found in K18.

The details of the \chandra\ analysis to obtain the halo image are given in K18, except for one particular difference. In K18, we used blank-sky images to determine the diffuse X-ray background in each band. The resulting profile showed a flat surface brightness up to 80\arcsec. However, none of our generated images show emission inside 80\arcsec, the $\Delta t$ values (see Eqn.~\ref{eqn:deltat}) corresponding to the scattering angles $<$80\arcsec are small, and the corresponding input flux is very low. However, as discussed in \S\ref{subsec:newdust}, some of the excess emission may be caused by double scattering, which is not negligible. The blank-sky method possibly underestimates the background in surface brightness. Therefore, in Chip 7, we picked an annular region between 30\arcsec\ and 60\arcsec\ around the source, and in Chip 6, a circular region of size 20\arcsec\ away from the halo to determine new background figures in each chip. 

\subsection{X-ray spectral analysis and light curve}
\label{subsec:xray}

 Another key difference between K18 and this work lies in the treatment of X-ray flux that interacts with molecular clouds. Unlike K18, where clouds are modeled as homogeneous spheres, this study adopts a more detailed approach, requiring a precise determination of extinction per pixel and scattering angle (see Eqn.~\ref{eqn:extinction} and Fig.~\ref{fig:dshgeom}). This means computing the unabsorbed flux from the source that passes through the molecular clouds, accounting for both scattering and absorption effects more rigorously.

We used two sets of data to achieve our goal: the \maxi\ 2--4 keV light curve, which covers the entire 2016 outburst; and pointed \swift\ XRT observations, covering both the first half of the outburst in the windowed timing (WT) mode (MJD 57630.5--57682.5, see Table~\ref{table:obs}) and the decay part (MJD 57769.8--57777.4), mostly in the Photon Counting (PC) mode (see K18). We used High Energy Astrophysics Software (HEASOFT) v6.31 and produced photon event lists
and exposure maps using $xrtpipeline$. For both WT and PC observations, we extracted photons from a circular region with a 20-pixel (47\arcsec) radius centered on the source. For WT mode observations, the background region is an annulus between 90-110 pixels (211.5\arcsec--258.5\arcsec). An appropriate background scaling is applied. The PC spectra for the observations during the decay have already been produced as described in K18.

The unabsorbed light curve estimation is done in three distinct regions: (1) MJD 57625--57682 where both \maxi\ and \swift\ XRT observations are present, (2) MJD 57683--57753 where only \maxi\ data are present, and (3) MJD 57754--57789, which is the decay part with \swift\ XRT observations reaching to the \chandra\ observation. In this region, the \maxi\ light curve is not reliable, possibly due to the nearby bright source GX~340$+$0 affecting the \maxi\ background. 

 We first extracted spectra from all \swift\ observations in region (1) and fitted them with an absorbed diskbb model ($tbabs \times diskbb$ with $wilm$ abundances \citep{Wilms00}). No other component was required in the fits. We then obtained 2--4 keV fluxes. By comparing those with the \maxi\ fluxes in the same energy band, we obtained a cross-calibration factor. We also obtained unabsorbed light curves in 1.5--2.25 keV (E1), 2.25--3.15 keV (E2), and 3.15--5 keV (E3) using the same \swift\ observations. Since there was minimal spectral evolution and no sign of state transition throughout the flat part of the outburst (see Table~\ref{table:obs}), we obtained a scaling factor by dividing the average unabsorbed \swift\ fluxes by the average \maxi\ 2--4 keV flux.

 In region (2), we simply multiplied \maxi\ 2--4 keV fluxes with the scaling and cross-calibration factors obtained in region (1). This region ends at MJD 57753 beyond which the \maxi\ background becomes unreliable.
 
\begin{figure}
\includegraphics[width=\columnwidth]{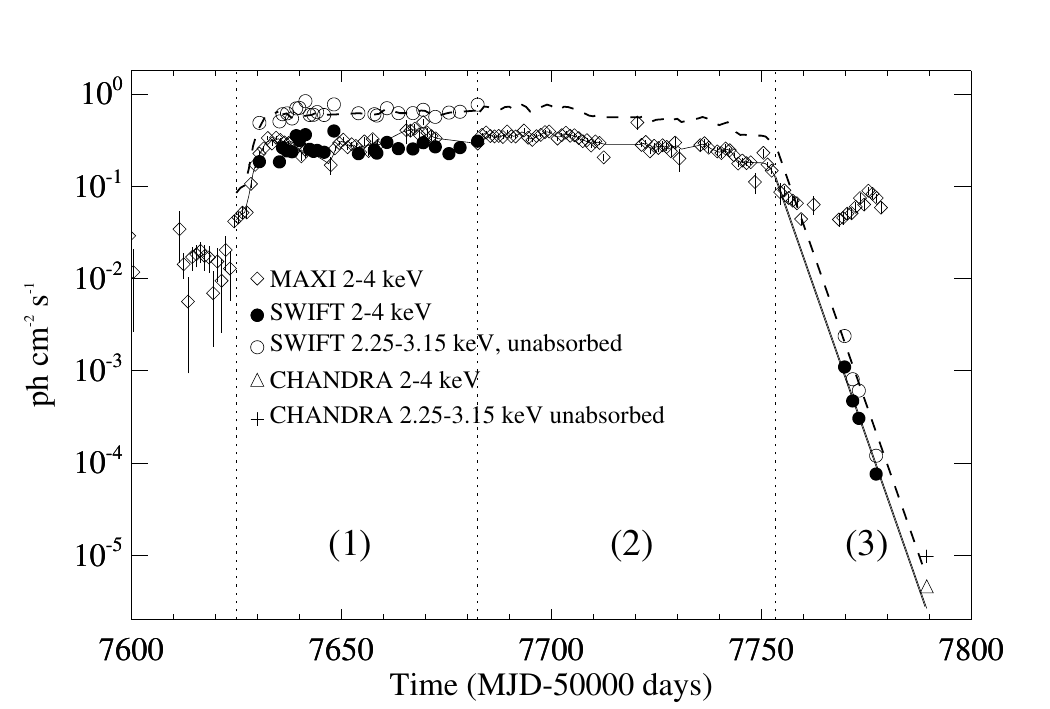}
\caption{\label{fig:lightcurve}
\maxi\ 2--4 keV and constructed unabsorbed 2.25--3.15 keV (E2) light curves. (1) to (3) denote regions in the light curve separated by vertical dashed lines. The solid line is a smoothed absorbed light curve, and the thick dashed-line is the smoothed unabsorbed 2.25--3.15 keV (E2) light curve. 
}
\vspace{0.8 pt}
\end{figure}

 In region (3), we fitted an exponentially decaying light curve in E1, E2, and E3 that passes through the unabsorbed fluxes of existing \swift\ observations to the \chandra\ unabsorbed flux.  For regions (1) and (2), we smoothed out sudden dips and peaks most probably caused by poor exposure and background subtraction in \maxi. The absorbed and unabsorbed light curves are shown in Fig.~\ref{fig:lightcurve}. Only E2 is shown for clarity.

\subsection{APEX data and analysis}

We performed spectral line mapping observations of the area around \FU{} using the \apex{} 12~m telescope \citep{Gusten2006} located on Chajnantor at an elevation of 5100~m with the 230~GHz channel of the nFLASH receiver. nFLASH is a dual-sideband receiver, providing $2 \times 8$GHz bandwidth with a $\sim 8$GHz gap in between. For our observations the receiver was tuned to cover frequencies between $\sim$229--237~GHz (in the upper side band -- USB) and between $\sim$213--221~GHz (in the lower side band -- LSB), including the frequencies of the $J = (2$--$1)$ main CO isotopologue lines: $^{12}$CO(2--1), $^{13}$CO(2--1), and C$^{18}$O(2--1) at rest-frequencies of 230.538000~GHz,  220.398684~GHz, and 219.560358~GHz, respectively. The spectra were recorded using the \apex{} FFT spectrometer at a spectral resolution of $\sim$244.1~kHz/$\sim$0.32~km/s in 2021 and at the native spectral resolution of $\sim$61~kHz/$\sim$0.079~km/s in 2022.

Observations were taken on 2021, May 08, and in 2022 on July 09/10/11 and on October 11, all in good to very good weather conditions (precipitable water vapor between 0.8~mm and 3.4~mm). In 2021, a 5$\times$5\amin{} map centered on the position of \FU{} was taken, while in 2022 a larger 6$\times$9\amin{} region (oriented N--S with a position angle of 3.4\degree{} west--of--north) was observed. Mapping was done in the On--The--Fly (OTF) mode, continuously scanning across the map region while dumping spectra at positions spaced by $\sim$1/3 of the beam width ($\sim$30\arcsec{}); correspondingly, spacings between scan rows were also set to $\sim$1/3 of the beam width. Several map coverages were required to reach the final exposure, where the scanning direction was alternated between east--west and north--south in order to reduce scanning artifacts. Atmospheric calibrations were interspersed every few minutes, while pointing checks were done every 1--2~hours.

Data reduction followed standard procedures within the GILDAS/CLASS software\footnote{see https://www.iram.fr/IRAMFR/GILDAS}. Low-order polynomial baselines were fitted and subtracted from the online-calibrated spectra, where spectral ranges showing clear line emission were excluded from the fits. Finally the spectra were gridded into cubes, one for each of the three lines, with a spectral resolution of 0.5~km/s and 500 channels spanning the velocity range from $-$175~km/s to $+$75~km/s. The spectra were then fitted with Gaussians to determine the central radial velocities of the molecular clouds (see Fig.~\ref{fig:NHIfit}, top two panels).

\subsection{SGPS analysis to obtain $N_{HI}$ distribution}
\label{subsec:NHI}

\begin{figure*}
\includegraphics[width=1.5\columnwidth]{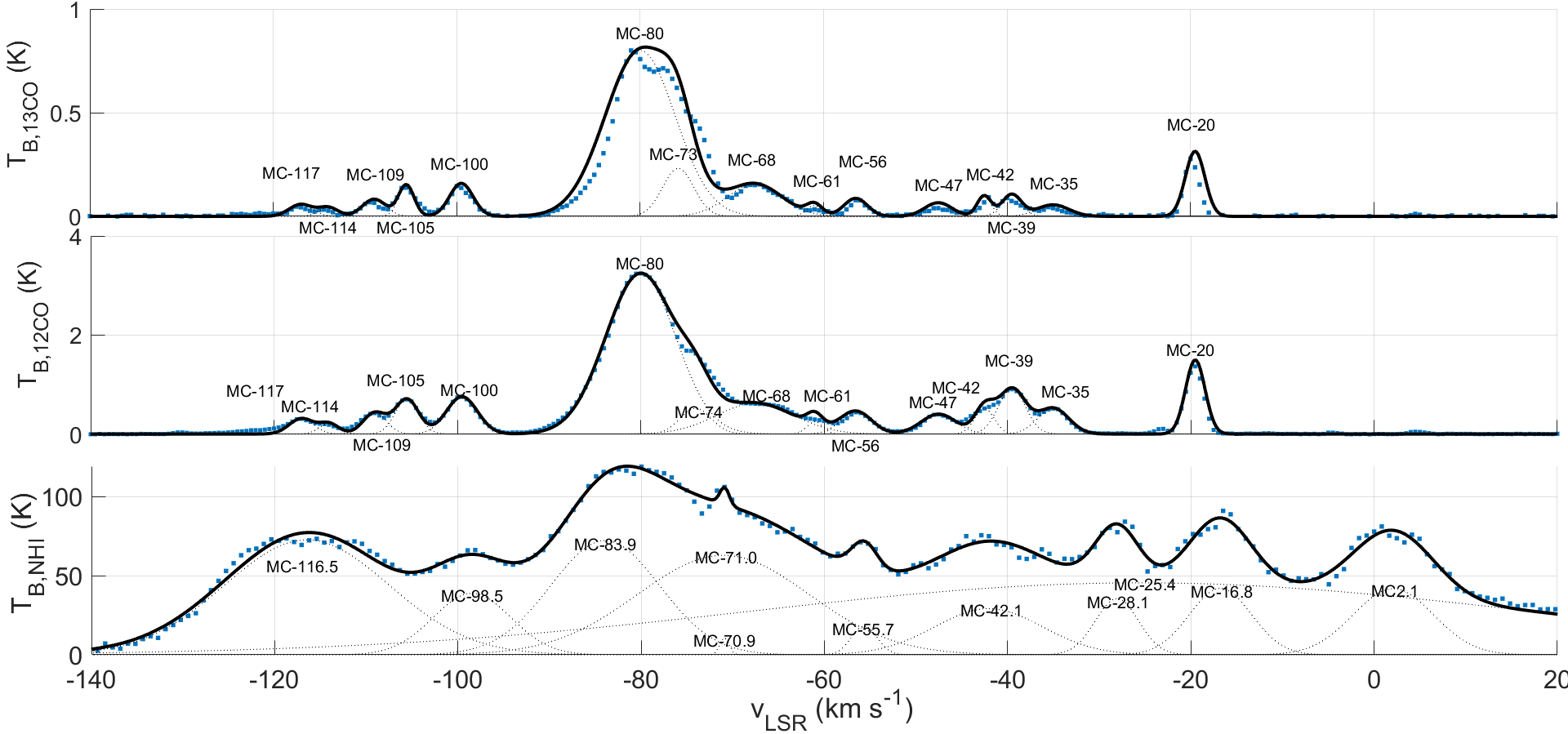}
\caption{\label{fig:NHIfit}
Top: $^{13}$CO spectrum from \apex. Middle: $^{12}$CO spectrum from \apex. Bottom: $N_{HI}$ spectrum from the Southern Galactic Plane Survey (SGPS) towards \fu. $T_{B,HI}$, $T_{B,12CO}$ and $T_{B,13CO}$ are the brightness temperatures of the emission lines. Gaussian decomposition is applied to all spectra.}
\vspace{1.5 pt}
\end{figure*}

We use publicly available data from the Southern Galactic Plane Survey (SGPS) to estimate the total column density corresponding to neutral atomic hydrogen along the line of sight to \fu. The spectral data can be decomposed into 11 Gaussian components, spanning radial velocities from --150 km s$^{-1}$ to ~50 km s$^{-1}$ (Fig.~\ref{fig:NHIfit}, bottom panel). To find the total column density corresponding to neutral atomic hydrogen along the line of sight to the source, we can integrate the brightness temperature over the velocity range as follows \citep{Draine2011}:
\begin{equation}
    N_{HI}=1.823\times10^{18}\int T_{B,HI}(v_{LSR})dv \; \rm{cm^{-2}}
    \label{eq:NHI}
\end{equation}

Note that this holds for the assumption of optically thin emission. For denser areas of \HI, this provides a lower limit since self-absorption of the \HI\ 21cm line can occur \citep{HI4PI}. Integrating over the whole velocity spectrum via Eqn.~\ref{eq:NHI} yields a total neutral atomic hydrogen column density of $2\times10^{22}$ cm$^{-2}$, which agrees with the lower limit found by \cite{Augusteijn01}. 

\begin{figure}
\includegraphics[width=\columnwidth]{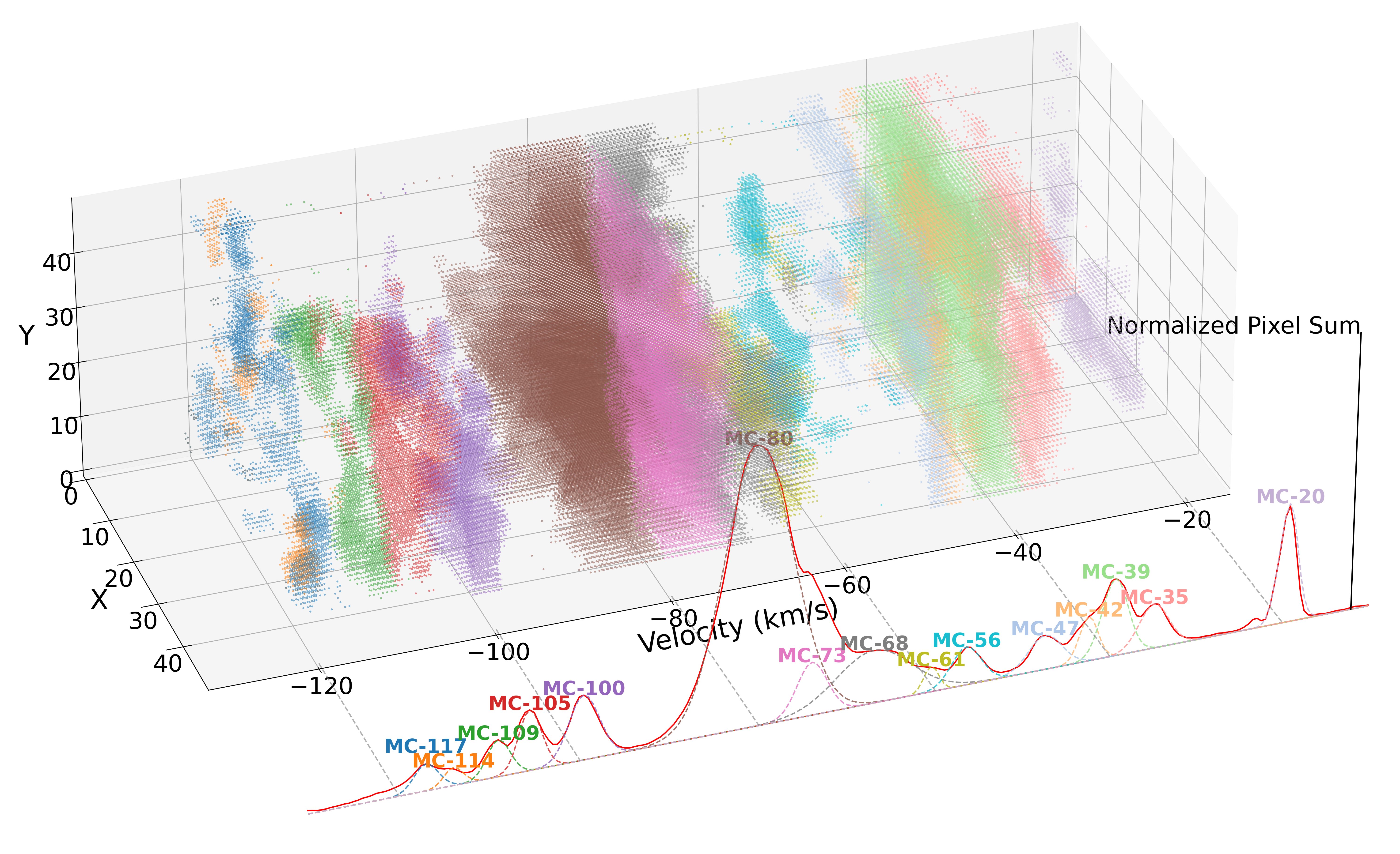}
\caption{\label{fig:3drep}
    The foreground shows the velocity spectrum with fitted Gaussians, while the background displays the 3D shapes of molecular clouds based on x, y, and velocity coordinates. The colors of the Gaussians and 3D shapes are matched.}
\end{figure}

\begin{figure}
\includegraphics[width=\columnwidth]{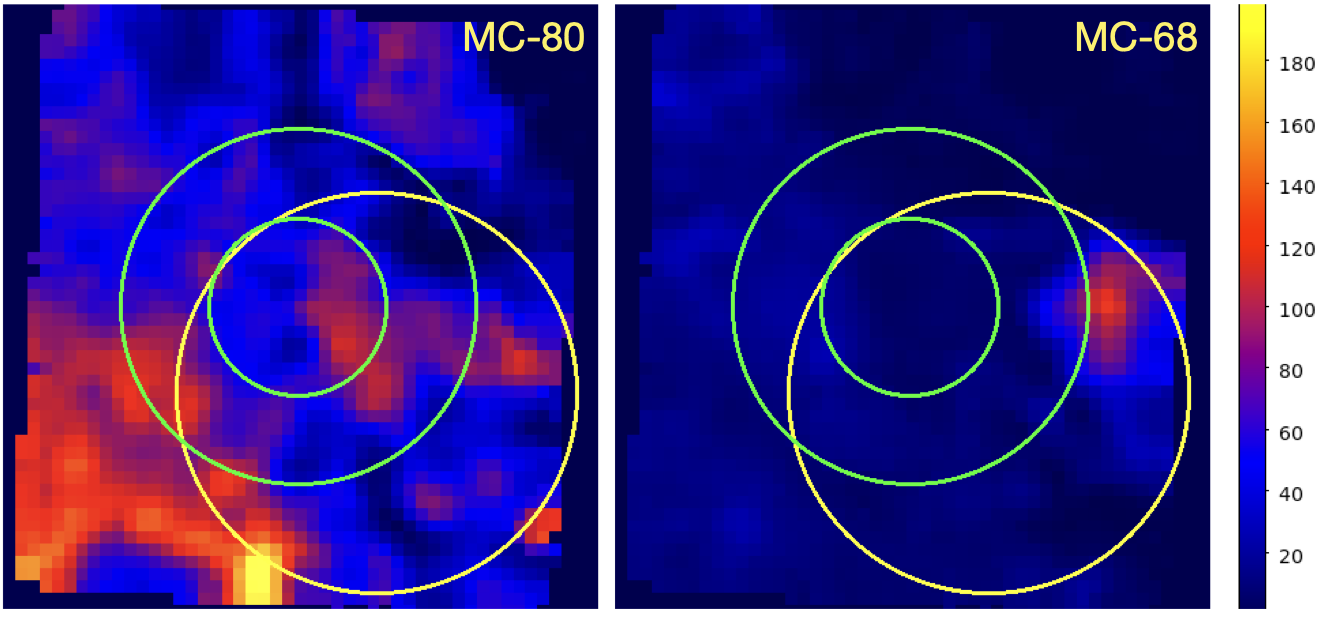}
\caption{\label{fig:apex_sample}
\apex\ integrated cloud images for MC-80 (left) and MC-68 (right). The green annulus shows the position of the DSH ring. The yellow circle is the Bronfman Survey pointing.}
\end{figure}

\subsection{Determining cloud shapes}
\label{subsec:detshapes}
 
We used millimeter data from 500 high-resolution \apex\ velocity channel maps to determine the 3D representations and densities of molecular clouds along the line of sight. Because of the small field of view and overlapping data, we developed a novel set of algorithms and test functions to determine the cloud shapes as explained in detail in the Appendix, and here we provide a brief summary. The analysis starts with Gaussian decomposition to identify peak radial velocities and velocity dispersions of molecular clouds from the $^{12}$CO brightness spectra. After the images are segmented into velocity groups, thresholding is applied to isolate cloud regions. Finally, a modified Mean-Shift clustering algorithm groups data points into 15 distinct clouds and reconstructs their 3D representations with densities (see Fig.~\ref{fig:3drep}). We name the molecular clouds with peak radial velocities such as ``MC-80". In Fig.~\ref{fig:apex_sample}, we show two representative \apex\ images, obtained through integrating brightness over the velocity slices of the given clouds to highlight the advantage of having high-resolution images. These images also show the area of the main DSH ring, as well as the low-resolution Bronfman Survey \citep{Bronfman89} pointing. In K18, MC-68 was particularly discussed as it was utilized to fit the profile for radii smaller than 80\arcsec. There was also some discussion about whether the source is inside this cloud. With the high resolution image, we can see that MC-68, while relatively bright, only covers a small area with a small intersection with the main ring; therefore, the profile at small radii is probably background dominated (see \S\ref{subsec:xray}), and there is no strong presence of this cloud at the source position. 

\begin{table}
\caption{Cloud Parameters}
\centering
\label{table:cloudpar}
\begin{tabular}{cccccc}
\hline
No & \textbf{Name} & \textbf{Near} & \textbf{Far} & \textbf{W(CO)$^{a}$} & \textbf{N$_H^{b}$} \\
  &  & \textbf{(kpc)} & \textbf{(kpc)} & \textbf{(K km s$^{-1}$)} &$\bf 10^{22} \, \rm \bf cm^{-2}$\\
\hline \hline
1 & MC-20  & 1.82 & 13.87 & 3.66 & 0.15\\ 
2 & MC-35  & 2.71 & 12.91 & 1.91 & 0.076 \\ 
3 & MC-39  & 2.92 & 12.68 & 4.39 & 0.18 \\ 
4 & MC-42  & 3.07 & 12.52 & 1.10 & 0.044 \\ 
5 & MC-47  & 3.31 & 12.26 & 1.28 & 0.051\\ 
6 & MC-56  & 3.72 & 11.83 & 1.46 & 0.058\\ 
7 & MC-61  & 3.93 & 11.60 & 1.38 & 0.055\\ 
8 & MC-68  & 4.21 & 11.31 & 6.98 & 0.28\\ 
9 & MC-73  & 4.40 & 11.10 & 7.07 & 0.28\\ 
10 & MC-80  & 4.67 & 10.83 & 29.49 & 1.18\\ 
11 & MC-100 & 5.38 & 10.09  & 2.99 & 0.12\\ 
12 & MC-105 & 5.55 & 9.91  & 2.49 & 0.099\\ 
13 & MC-109 & 5.69 & 9.76  & 1.19 & 0.048 \\ 
14 & MC-114 & 5.87 & 9.58  & 0.72 & 0.029 \\ 
15 & MC-117 & 5.98 & 9.47  & 0.94 & 0.038\\ \hline
\end{tabular}

\small
$^{a}$ Integrated $^{12}$CO emission ($J = 1-0$)\\
$^{b}$Estimated $N_{H}$ from the Bolatto relation \citep{Bolatto13}

\end{table}

Each centroid radial velocity corresponds to a near and far distance for each cloud. These distances are ascertained using the Galactic rotational curve model used by \cite{Reid14} similar to K18. Table~\ref{table:cloudpar} summarizes the decomposition of $^{12}$CO data into molecular clouds. The values for W(CO) are found by integrating the average temperature in each slice of each cloud. We note that the W(CO) values are lower than for those reported in K18 for the same clouds. We found that intrinsically, the \apex\ brightness temperatures are lower than those obtained from the Bronfman data (they are also not covering exactly the same area as shown in Fig.~\ref{fig:apex_sample}). Moreover the thresholding removes an additional 10-15\%.

\subsection{DSH image generation}
\label{subsec:imgen}
To generate each DSH image for any given source distance, the following algorithm is used:
\begin{enumerate}
    \item Generate the permutation of near and far distances for all clouds, then sort them to consider only the ones in front of the source along the line of sight to the observer. This creates a 3D voxel array structure spanning the entire observational path.

    \item  Estimate total $N_H$ for each voxel.

    For each voxel, the total hydrogen column density $N_H$ is calculated by combining the molecular and neutral atomic hydrogen components. For the molecular contribution, we use $N_{H_{2}}(x_p,y_p,z) = X_{CO} \; T_{B,12CO}(x_p,y_p,z) \; 0.5 \; \rm km \, s^{-1}$ with the the conversion factor $X_{CO}= 2 \times 10^{20}$ \rm cm$^{-2} \; (\rm K\, km\, s^{-1})^{-1} $ \citep{Bolatto13}. $x_p$, $y_p$, and $z$ denotes the x, y positions of the scattering voxel in the \apex\ slice $z$, respectively.
    
    The neutral atomic hydrogen column density is split into two parts: a voxel-dependent component $N_{H,W}(x_p,y_p,z)$ proportional to the $N_{H_{2}}(x_p,y_p,z)$, representing the atomic hydrogen associated with the molecular clouds, and a uniform component $N_{H,U}(z)$ distributed evenly along the line of sight between 2.35--12.9 kpc. The second component corresponds to the neutral atomic hydrogen associated with the spiral arms, and the distances indicate the start and end of the spiral arms in the direction of \fu. In summary, the total hydrogen column density for each pixel can be written as
    
    \begin{equation}
        \label{eqn:NH}
        N_{H}(x_{p},y_{p},z)=2 N_{H_{2}}(x_p,y_p,z)  + N_{H,W}(x_p,y_p,z) + N_{H,U}(z)
    \end{equation} 
 
 See \S\ref{subsec:validity} for a discussion of the choice of $N_{H}$ distribution.

   \item Model the extinction of photons in two stages: from the source to the scattering voxel, and from the scattering pixel to the observer.
   
   The extinction coefficient for the pixel $(x_p,y_p)$ at slice $z$ is:

        \begin{equation}
         \label{eqn:extinction}
        K_z(x_{p},y_{p})= \left( \sum\limits_{z,as} N_{H}(x_p,y_p,z)+\sum\limits_{z,bs} N_{H}(x_a,y_a,z)\right) \times \sigma_{ph,E}
    \end{equation}
    
    where $\sigma_{ph,E}$ is the energy-dependent absorption cross-section, and  $x_a, y_a$ denote the positions of the absorbing voxels at each slice before scattering. The integrations follow the three-dimensional photon path through the dust and gas distribution. The first sum considers the extinction undergone by the photons from the scattering pixel to the observer ($z,as$: sum over slices after scattering). Since, after scattering, the photons propagate along the same direction used to generate the projected image (see Fig.~\ref{fig:dshgeom}), the $x_p$ and $y_p$ indices in each slice remain identical to those in the projected image. On the other hand, the second summation accounts for the extinction experienced by the photons along their trajectory from the source to the scattering pixel ($z,bs$: sum over slices before scattering). The pixel indices corresponding to each slice along this path are determined using spherical geometry, based on the distance of the slice and the distance to the source. To determine $\sigma_{ph,E}$, we utilized X-ray spectral fits and compared absorbed and unabsorbed fluxes in each energy band. 

\begin{figure}
\includegraphics[width=\columnwidth]{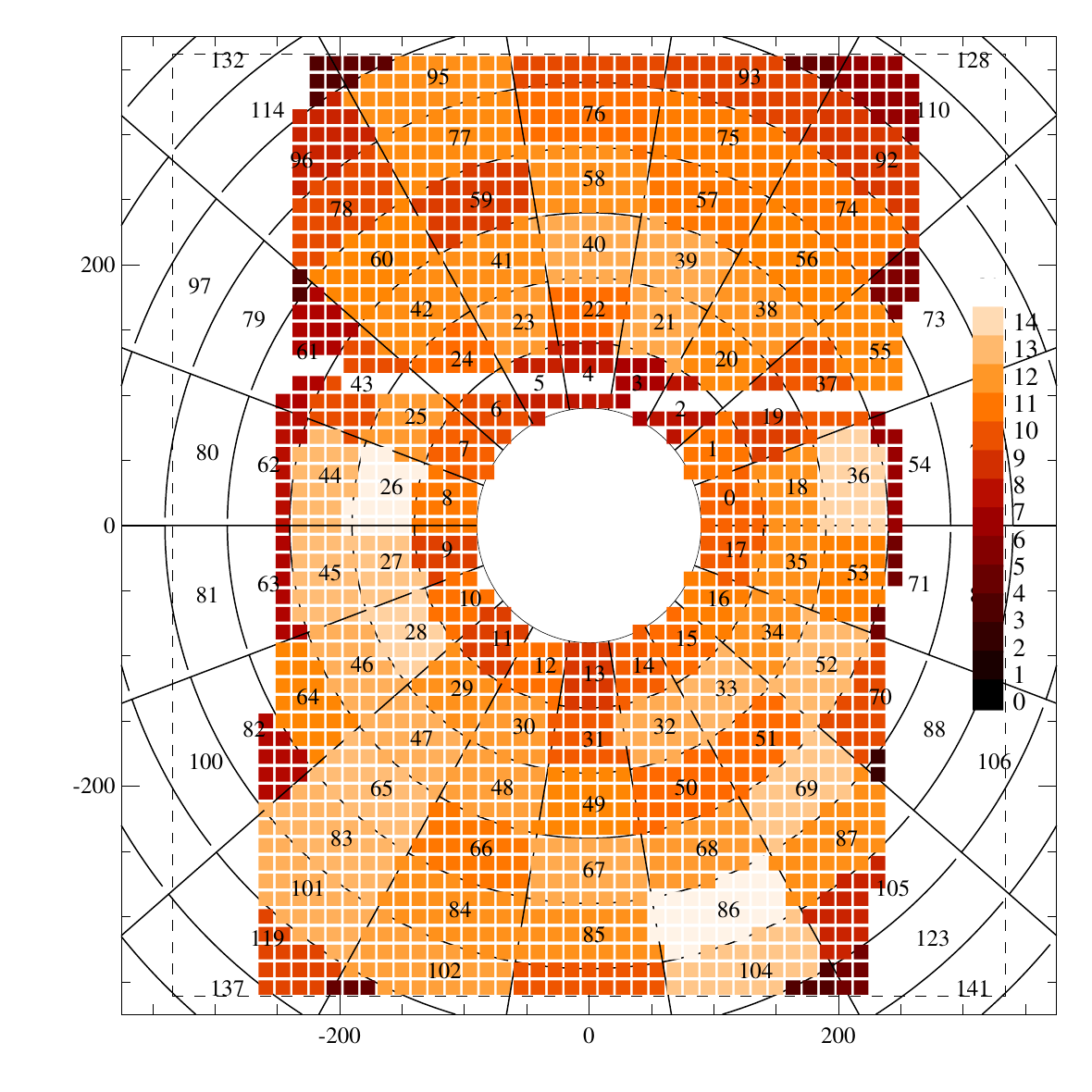}
\caption{\label{fig:wedges}
The \chandra\ and generated images are segmented with the wedge-shaped bins as shown here. The dashed lines roughly represent the  \apex\ field of view. The colors represent the signal-to-noise ratio in \chandra\ E2 band in the given wedge. 
}

\end{figure}

    There may also be some local absorption at the source. Such absorption will enter the calculation as $e^{-N_{H, local} \sigma_{ph,E}}$ for all pixels; hence it just changes normalization. See Section~\ref{subsec:extinction} for a discussion of local absorption.

    \item Determine the time delay of arrival of scattered photons, $\Delta t$:
   \begin{equation}
       \label{eqn:deltat}
	\Delta t = \frac{xD\theta^{2}(x_{p}, y_{p})} {2c(1-x)}
    \end{equation}
    \\
where $x$ is the ratio between the voxel distance $d$ and source distance $D$ ($x=d/D$), $c$ is the speed of light (see Figure~\ref{fig:dshgeom}).
    
\item Determine dust scattering halo intensity $I_{\nu} (x_{p}, y_{p})$  by integrating input flux as a function of time and scattering cross section per Hydrogen column density over all slices while taking absorption into account:
    \begin{equation}
    \begin{split}
	\label{eqn:flux}
	I_{\nu}(x_{p}, y_{p}) = \sum\limits_{z} N_{H}(x_p,y_p,z) \frac{d\sigma_{sc,E}}{d\Omega} \frac{F_{\nu} (t=t_{obs}-\Delta t)}{(1-x)^{2}} \times \\
    \exp{[-K_z(x_p,y_p)]}
    \end{split}
\end{equation}

The multiplicative term $N_{Hx_p,y_p,z}$ is determined via Eqn.\eqref{eqn:NH}.
The term 
 $\frac{d\sigma_{sc,E}}{d\Omega}\frac{1}{(1-x)^{2}}$
is found for each $x$ and $\theta$ using the latest version of the \textit{NewDust} code \citep{Corrales23}. This term corresponds to the $norm\_int$ in \textit{NewDust}, which is the surface brightness divided by the X-ray flux. \textit{NewDust} uses a custom-defined grid of energy and $\theta$ as inputs. See Section~\ref{subsec:newdust} for a discussion of the \textit{NewDust} cross sections. We chose Rayleigh-Gans dust model with Drude approximation \citep{Corrales16}, and an $md$ parameter (dust mass column in $\mathrm{g} \, \mathrm{cm}^{-2}$) of 0.000232 per $\rm 10^{22} \; cm^{-2}$ of $N_{H}$ which corresponds to a dust-to-mass ratio of 0.01 (L. Corrales, personal communication). The $norm\_int$ values are computed for $\theta$ ranging from 0\arcsec\ to 600\arcsec\ with $x$ values ranging from 0.001 to 0.999 with 0.001 resolution for 3 energy bands (E1, E2, E3).

    \item Add the World Coordinate System (WCS) and save the generated image as a FITS file.
    
\end{enumerate}

\begin{figure*}
\centering
\includegraphics[width=.29\linewidth]{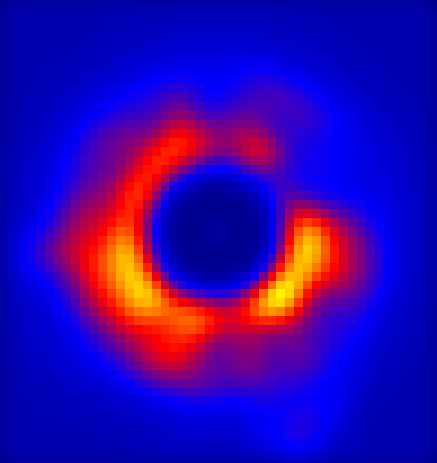}
\includegraphics[width=.47\linewidth]{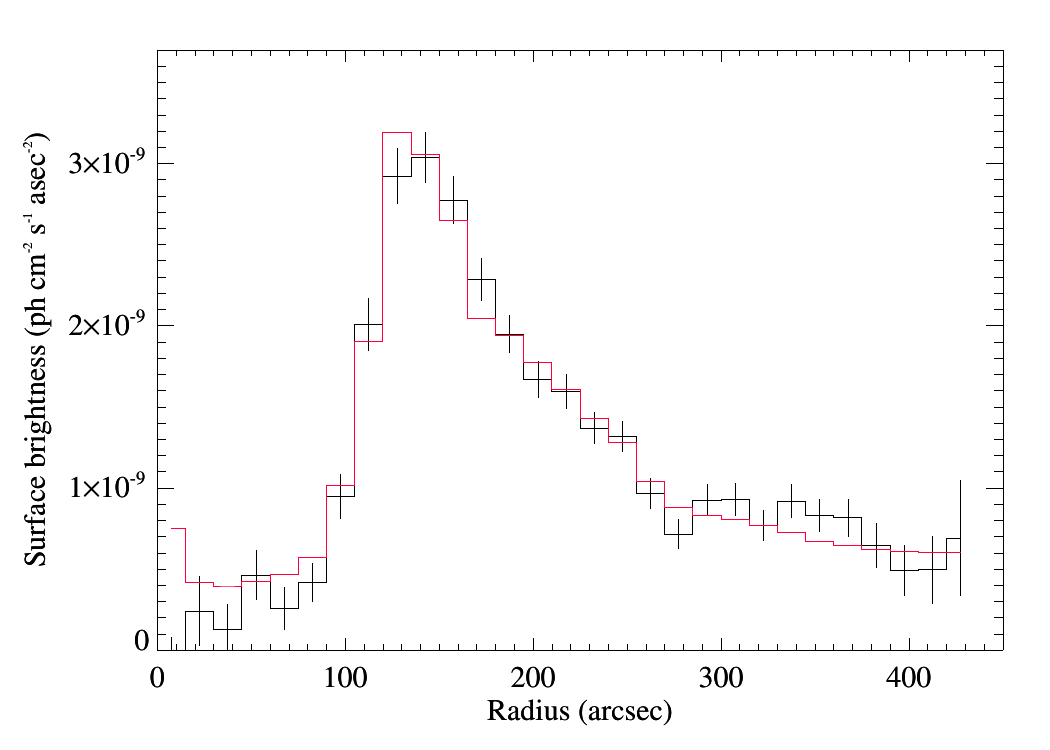}
\includegraphics[width=.29\linewidth]{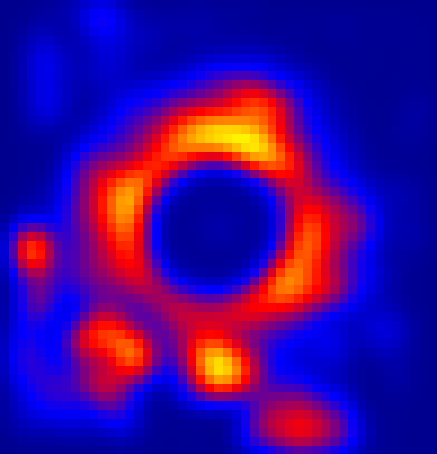}
\includegraphics[width=.47\linewidth]{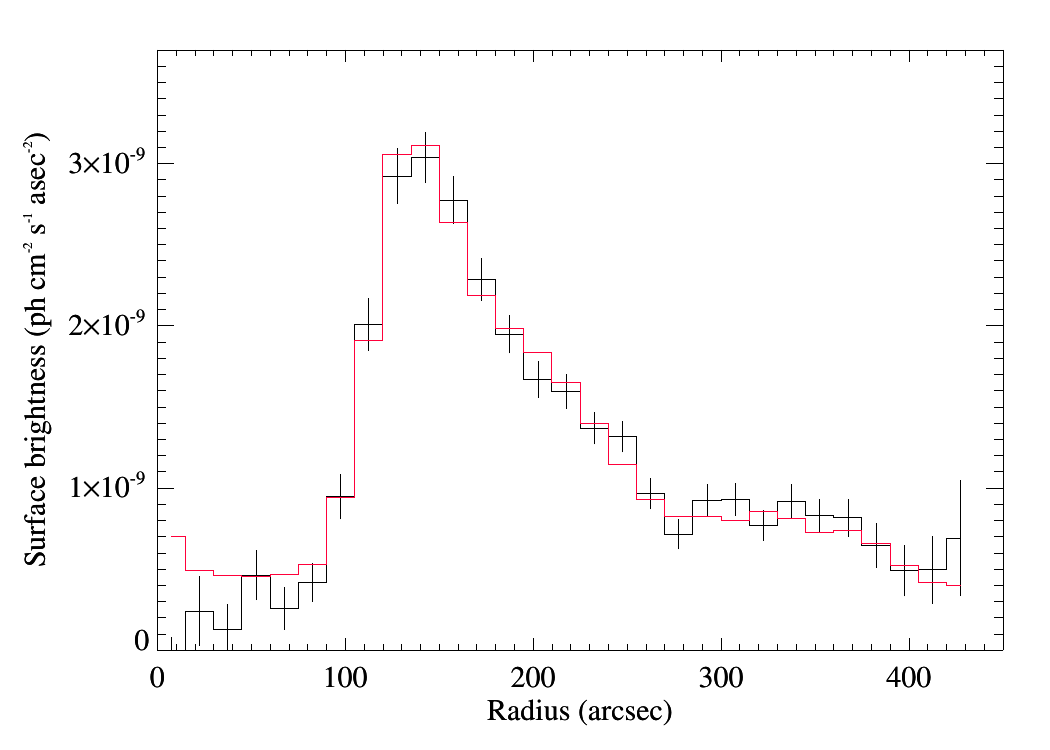}
\caption{\label{fig:radfits}
Top: The best fit generated image and radial profile fit to the \chandra\ profile in E2 for 11.5 kpc with $N_{H,U} = 2.5 \times 10^{22} \, \rm{cm}^{-2}$ with no $N_{H,W}$. Bottom: Same as top, but for a distance of 13.6 kpc with no $N_{H,U}$ or $N_{H,W}$.
}
\vspace{0.5 pt}
\end{figure*}

\subsection{Generated image post-processing}
\label{subsec:postpro}

To compare the generated images with the \chandra\ image, we first removed detected point sources from the \chandra\ images in the given energy bands as described in K18. We then rebinned the \chandra\ images to match the resolution of the \apex\ images. We determined pixels with low exposures ($<$ 50\% of the maximum, mostly corresponding to the chip boundaries) using the \chandra\ exposure map, flagged and excluded them from the analysis. Fig.~\ref{fig:chrebine2} is the resulting image in the E2 band we used for direct comparison with the generated images.

We created radial surface brightness profiles (SBP) to compare with our previous results in K18. We summed photon fluxes inside radial rings of thickness 15\arcsec, and divided them by the angular area of each ring to obtain the SBP. The largest ring ends at 420\arcsec\ to cover the entire \apex\ field of view. The flagged pixels were removed from both the sum and the areas. The same procedure is applied to the generated images. The errors in the \chandra\ radial profile are calculated assuming Poisson statistics. We then fitted the \chandra\ profiles with the generated image profiles with two parameters: a normalization and a constant. While the normalization was free, the constant was allowed to vary between --0.3 $\times 10^{-9}$ -- 8 $\rm \times 10^{-9} \; ph \;cm^{-2} \;s^{-1} \; asec^{-2}$ as it represents the systematic error in determining the background (see \S\ref{subsec:chandra}). For the fitting process, we used $\chi^2$ minimization with the $MPFIT$ package in IDL \citep{Markwardt2009}.

\begin{figure}
\includegraphics[width=0.9\columnwidth]{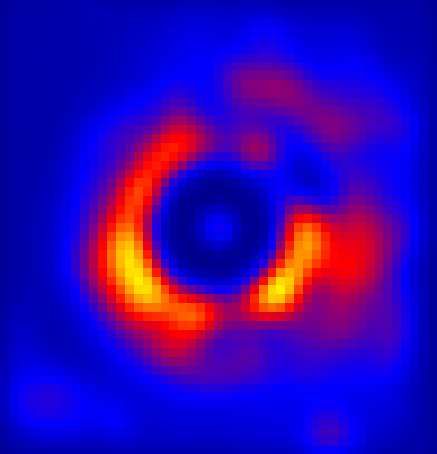}
\caption{\label{fig:best48rad}
The best fit image for 4.85 kpc distance. $N_{H,U} = 2 \times 10^{22} \rm{cm}^{-2}$ (left) with no $N_{H,W}$
}
\vspace{-1 pt}
\end{figure}

To take advantage of the high-resolution \apex\ images, we also produced azimuthal (or wedge) profiles. This is done by first dividing the radial region of interest into 8 rings of size 50\arcsec, and then dividing each ring into 18 azimuthal bins. After removing flagged pixels in \chandra\ images with low exposure, we calculated the signal-to-noise ratio (SNR) in each bin, again assuming Poisson statistics.  Finally, we made a cut on the SNR ($>$ 9) to obtain a ``wedge profile" with each wedge number corresponding to one of the bins in the image (see Fig.~\ref{fig:wedges}). We constructed the wedge profile for each generated image with the same bins, and again fitted, using normalization and a constant as free parameters (see \S\ref{subsec:wedge}). Note that the SNR does not always follow the brightness; the outer, larger wedges tend to have a larger SNR as the total counts are higher.

\begin{table*}
\caption{Radial Fit Results}
\centering
\label{table:radfits}
\begin{tabular}{ccccccc}
\hline
\textbf{Distance} & $N_{H,U}$ & $N_{H,W}$ &  \textbf{$\chi^2$} & \textbf{Norm} & \textbf{Bkg} & \textbf{$N_{H,s}$}\\
\textbf(kpc) & $10^{22} \, \rm cm^{-2}$ & $10^{22} \, \rm cm^{-2}$ &  &  & \textbf{($\rm ph \, cm^{-2} \, s^{-2} \, asec^{-2}$)} & \textbf($10^{22} \, \rm cm^{-2}$) \\
\hline \hline
11.5 & 0 & 0 & 2.33 & 0.72 & $\rm 6.3 \times 10^{-10}$  & 2.43  \\ 
11.5 & 1 & 0 & 1.77 & 0.169 &  $\rm 4.8 \times 10^{-10}$ & 3.30  \\ 
11.5 & 1 & 2 & 1.96 & 0.098 & $\rm 5.4 \times 10^{-10}$  & 4.02  \\ 
11.5 & 2.5 & 0 & 1.29 & 0.168 & $\rm 3.3 \times 10^{-10}$ & 4.61  \\ 
11.5 & 2.5 & 2 & 1.63 & 0.099 & $\rm 4.6 \times 10^{-10}$  & 6.37  \\ 
13.6 & 0 & 0 & 1.00 & 0.767 & $\rm 4.0 \times 10^{-10}$  & 2.43  \\ 
13.6 & 1 & 0  & 5.91 & 0.517 & $\rm -0.3 \times 10^{-11}$& 3.43  \\ 

4.85 & 2 & 0  & 2.11 & 1.048 & $\rm 2.8 \times 10^{-10}$  & 2.51  \\ \hline
\end{tabular}
\end{table*}

\section{Results}\label{sec:results}

The main idea of this work is to generate DSH images using the 3D cloud shapes and compare them with the actual \chandra\ image. The images may be created for any source distance, $N_{H,U}$ that represents the cumulative $N_{HI}$ distributed uniformly to all pixels, and $N_{H,W}$ that represents the cumulative $N_{HI}$ distributed proportional to the molecular cloud brightness in each pixel (see \S\ref{subsec:imgen}). Given that there are 15 clouds, and each cloud can be in the near or far distance estimate, 32768 images need to be created and analyzed for each possible source distance, $N_{H,U}$ and $N_{H,W}$ combination. 

To reduce the number of images created, we used a simpler analytical model that assumes spherical clouds with the amount of dust proportional to $W(CO)$ (similar to the one used in K18), permuted source distances, and near- and far-choices for each cloud. This pre-analysis indicated that there are three possible distance ranges: 11.2--11.7 kpc, 13.4--13.8 kpc, and 4.7--5.2 kpc which would result in generated images resembling the \chandra\ image. Moreover, we further reduced the number of permutations using the same model; e.g. for the 11.2--11.7 kpc range MC-80 cannot be at the near distance as a bright ring would appear at  \wsim400\arcsec\ from the source, or for 13.4--13.6 kpc MC-80 cannot be at the far distance as we would have observed a double ring structure. We have generated images with all remaining available cloud permutations, 0.1 kpc distance steps and $N_{H,U}$ and $N_{H,W}$ at 0, 1, 2 and 2.5 $\times 10^{22}$ cm$^{-2}$ and applied fits described in \S~\ref{subsec:postpro}. For the 4.7--5.2 kpc range, we used 0.05 kpc distance steps to obtain a better fit. 

\subsection{Radial profile results}

For the 11.2 -- 11.7 kpc radial profiles, the minimum reduced $\chi^{2}$ is obtained for 11.5 kpc with uniform $N_{H,U}$ value of $2.5 \times 10^{22} \rm{cm}^{-2}$ with no $N_{H,W}$ (see Fig.~\ref{fig:radfits}). The reduced $\chi^{2}$ for the fit is 1.29 with a normalization of 0.168. The number of degrees of freedom for all radial fits is 24. The same figure also shows the best fit case for the 13.6 kpc distance with a reduced $\chi^{2}$ of 1.00. Notice the similar radial fits stemming from very different images. Both $N_{H,U}$ and $N_{H,W}$ are zero for the best 13.6 kpc case, the addition of any external $N_{H}$ makes the fit substantially worse as shown in Table~\ref{table:radfits}. In this table, $N_{H,s}$ is the average of total $N_{H}$ for the nine pixels encompassing the source position, calculated to compare with the $N_{H}$ values obtained from the X-ray spectral fits.

\begin{figure*} 
\centering
\includegraphics[width=.29\linewidth]{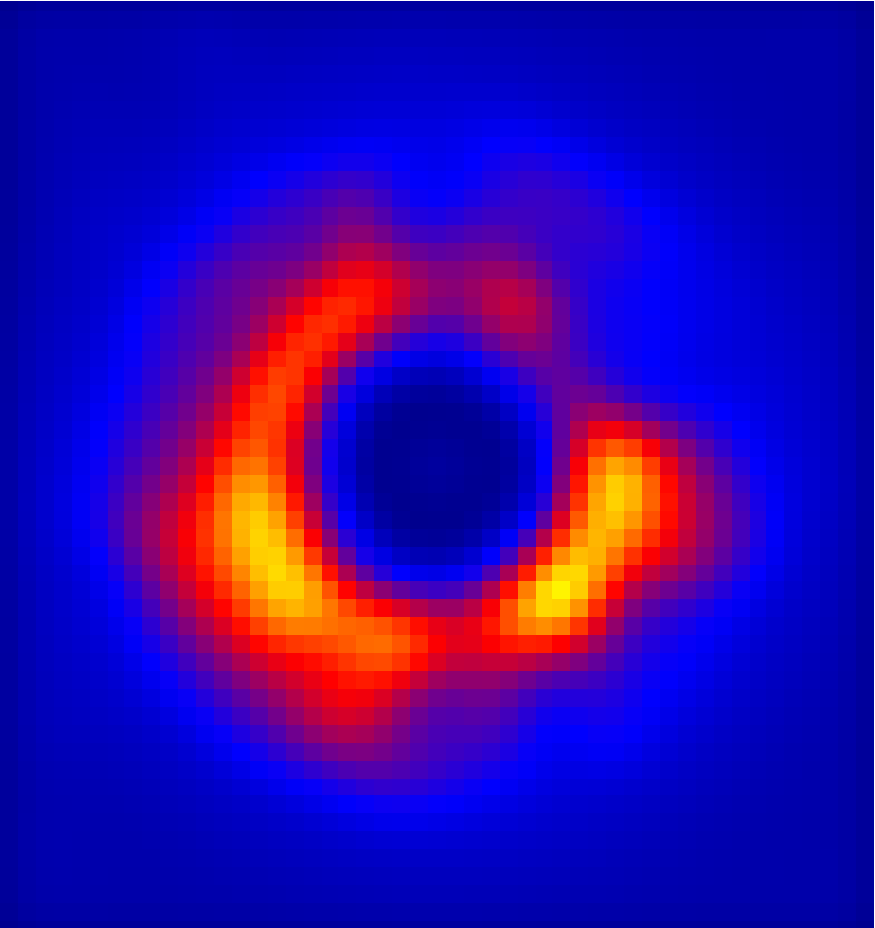}
\includegraphics[width=.47\linewidth]{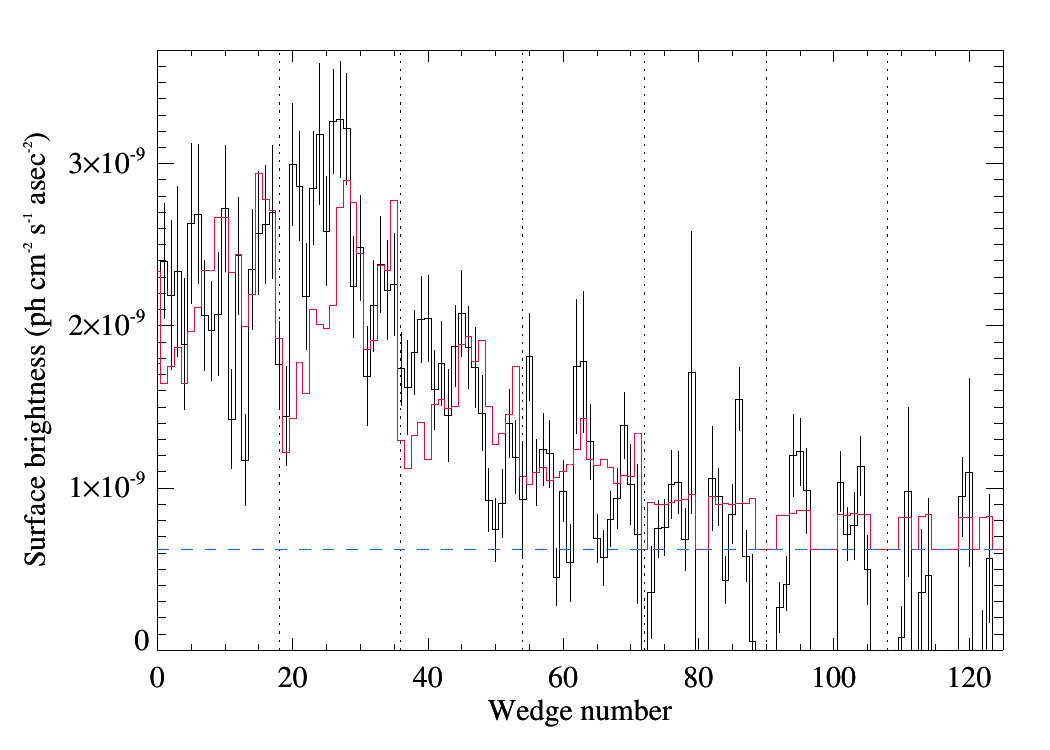}
\includegraphics[width=.29\linewidth]{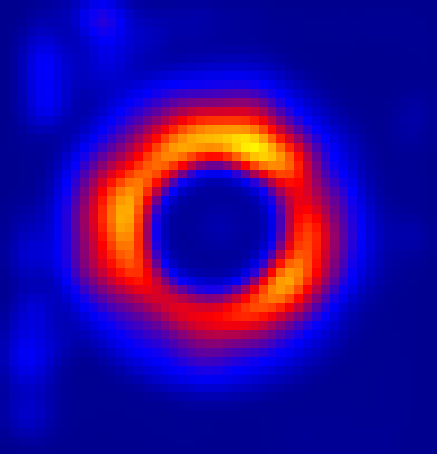}
\includegraphics[width=.47\linewidth]{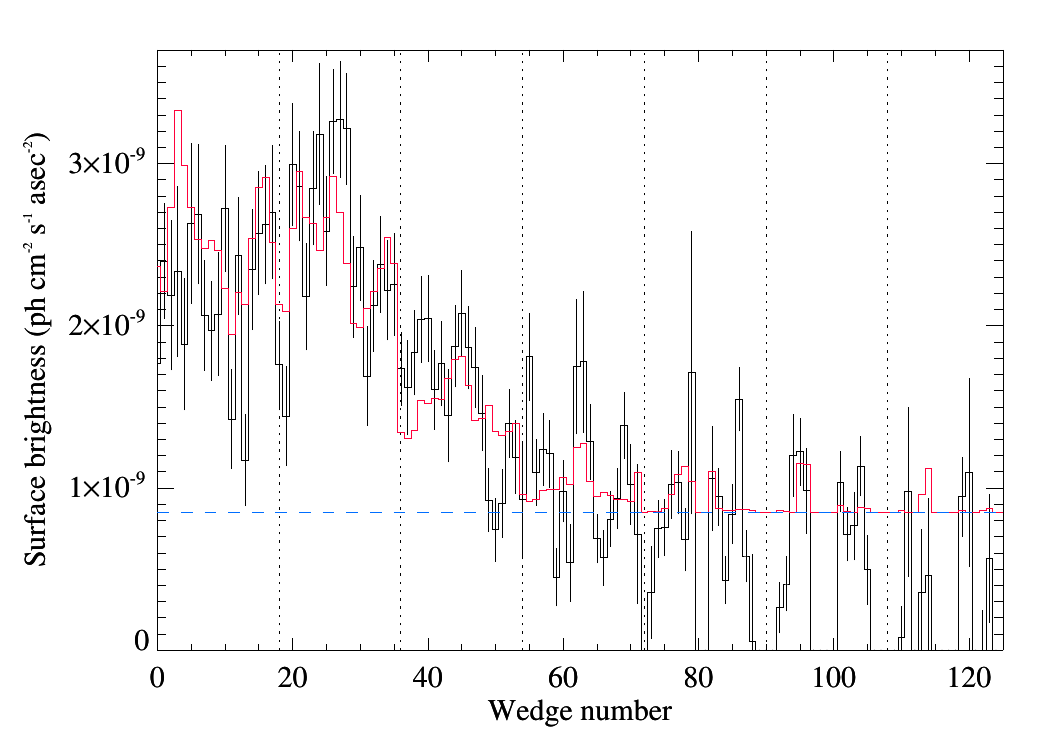}
\caption{\label{fig:wedfits}
Top: The best fit generated image and azimuthal profile fit to the \chandra\ profile in E2 for 11.5 kpc with $N_{H,U} = 2.5 \times 10^{22} \rm{cm}^{-2}$. Bottom: Same as top, but for a distance of 13.6 kpc and zero $N_{H,U}$ and $N_{H,W}$. See Fig.~\ref{fig:wedges} to match the wedge number to a region in the image. The vertical dotted lines limit radial boundaries, and the blue horizontal dashed line shows the best-fit background.
}
\vspace{0.5 pt}
\end{figure*}

\begin{table*}
\caption{Azimuthal Fit Results}
\centering
\label{table:wedgefits}
\begin{tabular}{ccccccc}
\hline
\textbf{Distance} & $N_{H,U}$ & $N_{H,W}$ &  \textbf{$\chi^2*$} & \textbf{Norm} & \textbf{Bkg} & \textbf{$N_{H,s}$}\\
(kpc) & $\rm 10^{22} \, cm^{-2}$ & $\rm 10^{22} \, cm^{-2}$ &  &  & ($\rm ph \, cm^{-2} \, s^{-2} \, asec^{-2}$) & ($10^{22} \, \rm cm^{-2}$) \\
\hline \hline
11.5 & 0-2.5 & 0-2 & 3.06-3.13 & 0.067-0.133 & $\rm 6.5-9.0 \, 10^{-10}$  & 2.43-6.37  \\ 
13.6 & 0 & 0 & 2.022 & 0.736 & $\rm 8.5 \, 10^{-10}$ & 2.43 \\
4.85 & 2 & 0  & 3.391 & 0.743 & $\rm 8.5 \, 10^{-10}$  &  2.511 \\ \hline
*Reduced $\chi^{2}$ with 67 degrees of freedom
\end{tabular}
\end{table*}

K18 near distance suggestion of 4.85 kpc (Fig.~\ref{fig:best48rad}) performs worse in radial profile fits compared to 11.5 kpc and 13.6 kpc distances as seen in Table~\ref{table:radfits}. It also indicated a low $N_{H}$ incompatible with the X-ray observations.

\subsection{Azimuthal profile results}
\label{subsec:wedge}

The azimuthal profile fits take full advantage of the high resolution \apex\ images and 3D cloud shapes. For the 11.2 -- 11.7 kpc range, the best fits are obtained for 11.5 kpc with very little change in reduced $\chi^{2}$ (3.06--3.13) when $N_{H,U}$ and $N_{H,W}$ were varied (see Table~\ref{table:wedgefits}). The best fit distribution and image are shown in Fig.~\ref{fig:wedfits}. The 13.6 kpc distance case yields a lower reduced $\chi^{2}$ of 2.02, which is also shown in Fig.~\ref{fig:wedfits}. Placing the MC-80 cloud in the near distance results in the best fit distance of 4.85 kpc with a reduced $\chi^{2}$ of 3.391.

\begin{figure}
\includegraphics[width=\columnwidth]{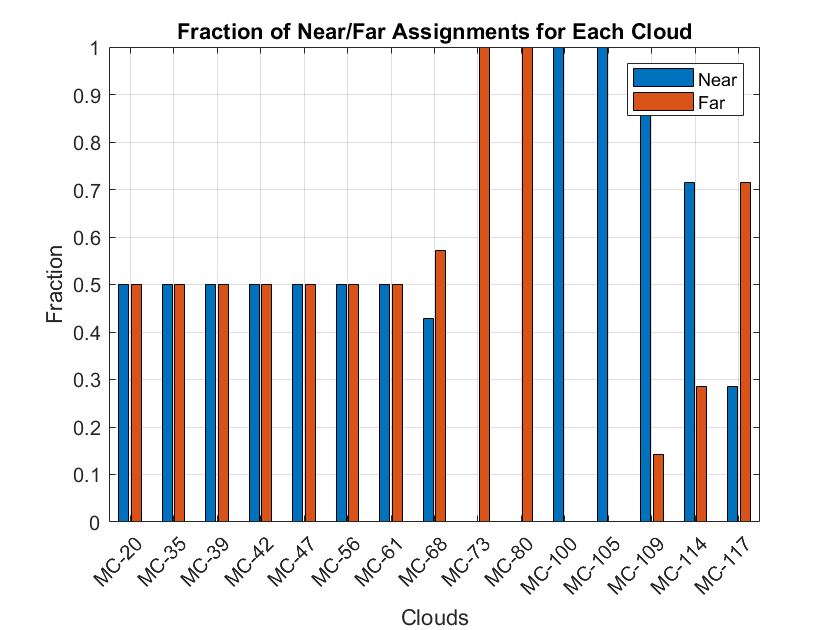}
\caption{\label{fig:wedcloudstats}
The fractions of clouds in near (N) and far (F) distances for 11.5 kpc with $N_{H,U} = 2.0 \times 10^{22} \rm{cm}^{-2}$ and no $N_{H,u}$ under reduced $\chi^{2}$ of 3.2 for azimuthal distribution fits.
}
\end{figure}

The azimuthal fits allow us to constrain the placement of clouds in the near- and far-distances. In Fig.~\ref{fig:wedcloudstats} we plot the fractional distribution of whether the clouds are in the near- or far- distance for all generated images with reduced $\chi^{2} <3.2$ for 11.5 kpc distance case. The $\chi^{2}$ cut corresponds to around 5\% of all generated images. The best fits require MC-73 and MC-80 to be at the far distance, and MC-100 and MC-105 at the near distance. The results also indicate that MC-109, and MC-114 are more likely to be at the near distance estimate, and MC-117 to be at the far distance estimate.

\subsection{Energy dependence}
\label{subsec:energydep}

While all the radial and azimuthal fits utilize the E2 band data, we performed an additional analysis using \chandra\ DSH images in the soft and hard bands (E1 and E3) in order to reveal the contributions of individual clouds to absorption and emission. By comparing the intensity distribution across these bands, we try to determine the order of clouds along the line of sight by correlating regions of higher absorption with cloud images, since clouds that exhibit higher absorption are likely positioned in the foreground. This approach, similar to that demonstrated by \cite{Heinz15}, not only clarifies the individual roles of the clouds but also helps in ruling out distance estimates that are inconsistent with the observed absorption and emission patterns.

\begin{figure}
\includegraphics[width=\columnwidth]{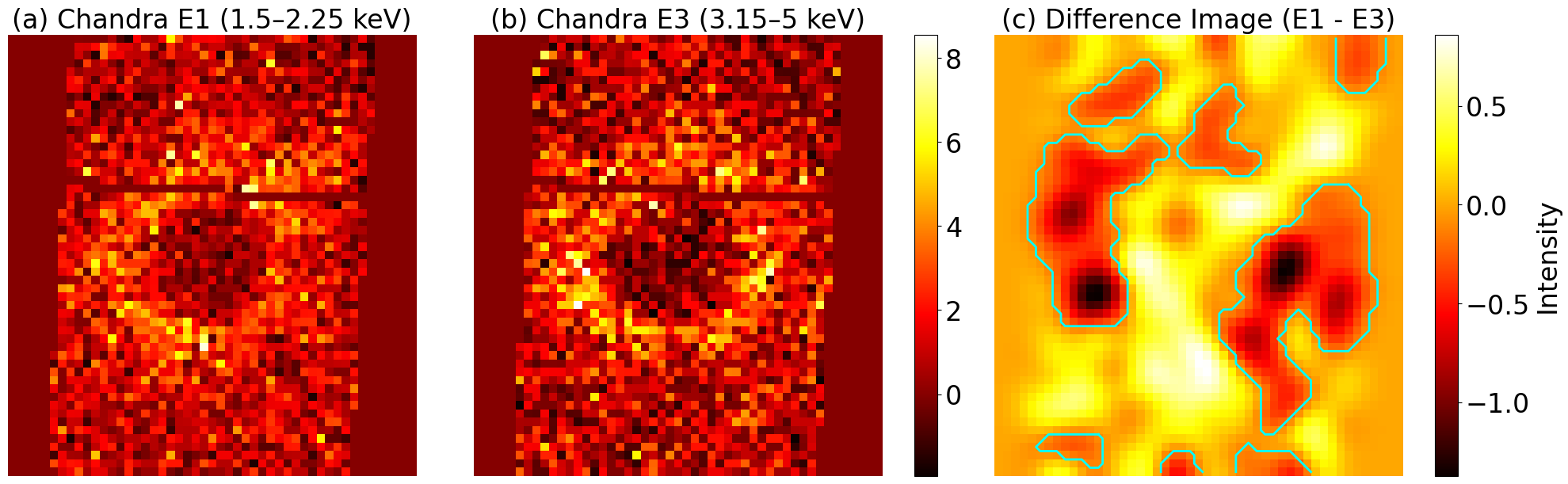}
\caption{\label{fig:energy_dep}
Normalized (with average brightness) \chandra\ images in E1 (a), E3 (b), and the smoothed difference (c). Highly absorbed regions are highlighted with contours.}
\vspace{0.3 pt}
\end{figure}

Firstly, we subtract the normalized E3 band \chandra\ image from that of E1 to highlight the absorbed regions (see Fig.~\ref{fig:energy_dep}).  Then, we integrate the 3D cloud velocity slices into one image for every cloud, using $^{13}$CO and $^{12}$CO data. Next, we compare the spatial overlap between the extracted difference image and each integrated cloud image. For this analysis, we generate a mask from the difference image by applying a threshold that isolates the regions with higher absorption. We then compute the overlap between these dark regions and the bright areas of the integrated cloud images to quantify the spatial correlation, thereby linking absorption features to specific molecular clouds.

Secondly, we executed our DSH simulations with the best-fitted parameters while preserving the full set of clouds along the line of sight and isolated the contribution of each cloud one by one to understand the individual contributions of the molecular clouds to the overall dust-scattering structure. This allowed us to identify clouds that are in front of the source and absorb X-rays but do not contribute strongly to the dust scattering image. We stress that our complete image-generation method takes these effects into account, therefore this is not an independent method, but it is much easier to implement without executing a large set of simulations.

\begin{figure}

\includegraphics[width=\columnwidth]{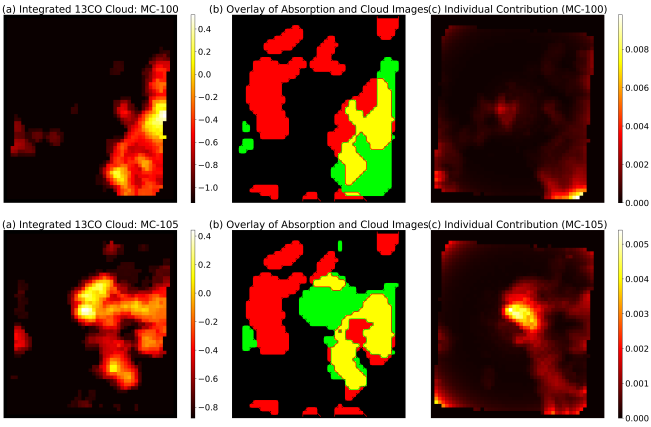}
\caption{\label{fig:MC-105_dep} (a)  The integrated $^{13}$CO image for the target clouds (MC‑100 and MC‑105); (b) an overlay where red indicates absorption regions, green indicates cloud regions, and yellow denotes their overlap, and (c) generated image isolating the individual contribution of the selected cloud (MC‑100 and MC‑105). The scale is with respect to the average of the full image shown in Fig.~\ref{fig:radfits}.
}
\vspace{0.3 pt}
\end{figure}

For a demonstration, we considered the highly absorbed region on the right side of the difference image. We found that MC-100 and MC-105 are likely to be responsible as they contribute mostly to the absorption in the given regions but not to emission (see Fig.~\ref{fig:MC-105_dep}). Since both near and far distance estimates of MC-100 and MC-105 are larger than 4.9 kpc, this independently rules out the 4.9 kpc case discussed in K18.

\section{Discussion}\label{sec:discussion}

Our methodology to generate DSH images using the high-resolution mm images from \apex\ includes a set of underlying assumptions, as well as limitations from the observing instruments. In this Section, we will start by discussing the implied distances from our results and then continue with the impact of our assumptions on our conclusions.  

\subsection{Is the source at 13.6 kpc?}
\label{subsec:thirteenpsix}

\begin{figure}
\includegraphics[width=\columnwidth]{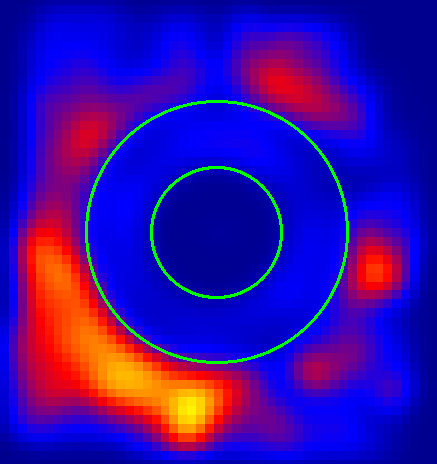}
\caption{\label{fig:135mc80far}
Generated image for the source in the 13.6 kpc case, and MC-80 being at the far distance. The green circles enclose the main ring observed in \chandra. 
}
\vspace{0.5 pt}
\end{figure}

Our method, for both the radial and azimuthal profile fits, indicates a lower $\chi^{2}$ value for a source distance of 13.6 kpc than that of 11.5 kpc. However, there are a couple of major problems with this distance estimate. The brightest molecular cloud, MC-80 cannot be at the far distance estimate for a source distance of 13.6 kpc, because it should produce a very bright second ring at 200\arcsec-400\arcsec that we do not observe (Fig.~\ref{fig:135mc80far}).

\begin{figure}
\includegraphics[width=\columnwidth]{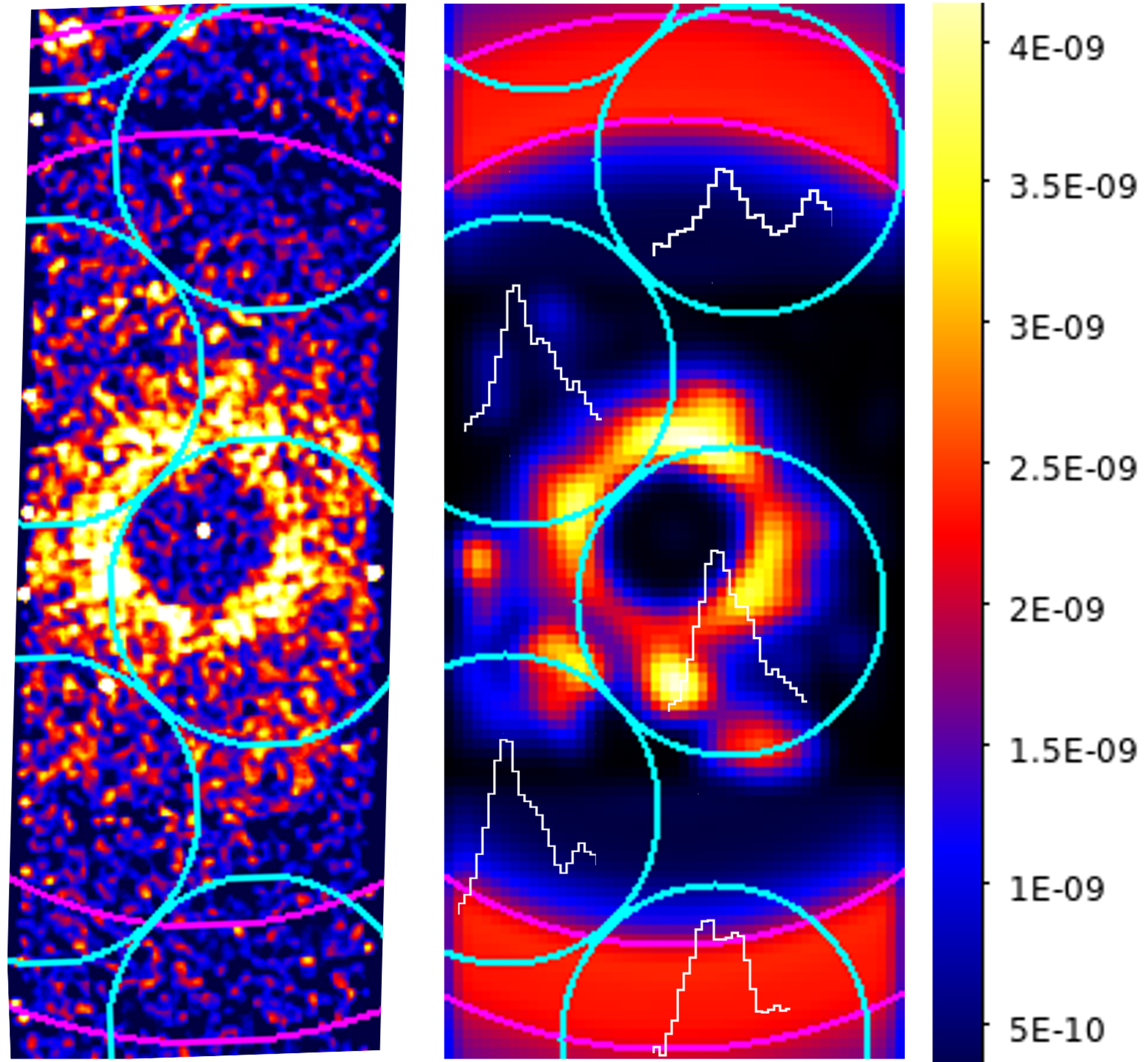}
\caption{\label{fig:bronfman} 
Left: \chandra\ E2 image with Bronfman Survey pointings overlaid with circular regions. Right: Generated image with source at 13.6 kpc and best radial fit parameters.  The $^{12}$CO spectra from the Bronfman Survey are also shown inside circles. The scale on the right is only for the generated image. The magenta annulus for both images shows the region between 600\arcsec\ and 750\arcsec.
}
\vspace{0.5 pt}
\end{figure}

The best fits indicate MC-80 to be at the near distance; however, our analysis does not extend beyond the field of view of \apex. From Bronfman Survey \citep{Bronfman89}, we know that MC-80 covers a much larger area than the \apex\ (see also K18 with a plot of --79 km/s cloud sizes). Assuming a similar and uniform $W(CO)$ beyond the \apex\ coverage, we generated a larger image encompassing the \chandra\ field of view. In Fig.~\ref{fig:bronfman}, one can see the \chandra\ E2 image and the generated image together. Bronfman survey pointings and the corresponding $^{12}$CO spectra are also overlaid. The generated image indicates a second ring between 600\arcsec\ and 750\arcsec. The average brightness in the outer ring of the generated image, 2.3 $\times$ 10$^{-9}$ ph cm$^{-2}$ s$^{-1}$ is almost the same as the average brightness between 100\arcsec\ and 200\arcsec\, where the observed ring is the brightest.  The \chandra\ image does not show any such bright outer ring. We note that there may be regions with less dust in MC-80 coinciding with the \chandra\ pointing, and indeed, the Bronfman Survey towards the North indicates a smaller peak at --79 km/s. However, the southern pointings are as bright as the central pointing, and yet there is no excess in the \chandra\ image.

The second reason for 13.6 kpc not being the source distance is the low total $N_{H}$ of $2.4 \times 10^{22}$ cm$^{-2}$ that the best fit indicates (see Table~\ref{table:wedgefits}). For the 13.6 kpc case, the main scatterers are MC-35 and MC-39 clouds, which possess much lower amounts of dust than MC-80 (see Table~\ref{table:cloudpar}). The fits do not allow adding $N_{H,W}$ or $N_{H,U}$. While we certainly are underestimating $N_{H}$ in general in the generated images (see \S~\ref{subsec:extinction}), it is not possible to make up for the missing column density for the case of 13.6 kpc source distance.

\subsection{Is the source at 4.85 kpc?}
\label{subsec:neardist}

Both radial and azimuthal fits result in a worse $\chi^{2}$ value for the 4.85 kpc case. We also showed in \S\ref{subsec:energydep} that the mm images indicate MC-105 and MC-100 as intervening clouds, yet both their near and far distances are larger than 4.85 kpc. Similarly to the 13.6 kpc case, the implied $N_H$ is significantly lower than values obtained from X-ray spectral fits.

In K18, the uniform $N_{H}$ was distributed between $x$ values of 0.1 and 0.99, even for the near distance case. This is likely to be unphysical for the near distance case, because all of the of the uniform \HI\ would be concentrated in less that 4.85 kpc from the solar system, and not in spiral arms. In this current work, we specifically distributed uniform \HI\ between 2.35 and 12.9 kpc and thereby include the spiral arms in the line of sight. We rule out the 4.85 kpc estimate more strongly than K18. It is also ruled out strongly through the analysis of X-ray polarization data by \emph{IXPE} \citep{Krawczynski23}.

\subsection{Is the source at 11.5 kpc?}

In K18, we selected MC-79 (MC-80 in this work) as the primary scattering cloud responsible for the main halo ring, since the Bronfman Survey identified it as the brightest CO emitter. This led to the conclusion that the source could be located either at 11.5 kpc or 4.9 kpc, depending on whether MC-79 lies at its far or near distance. However, a purely spatial analysis within the limited \apex\ field of view indicates that the radial and azimuthal halo distributions are better explained by scattering primarily in MC-35 and MC-39, which would instead place the source at 13.6 kpc. The extended \chandra\ image, however, rules out this scenario, as a strong secondary ring from scattering in MC-80 should have been present but is not observed.

By systematically excluding the 13.6 and 4.85 kpc distances (\S\ref{subsec:thirteenpsix} and \S\ref{subsec:neardist}, respectively), we conclude that 11.5 kpc is the most probable distance to the source.

\subsection{Validity of the approach and the distance error}
\label{subsec:validity}

The main hypothesis behind this study (and other similar works) is the presence of high-concentration dusty regions that scatter X-rays and produce the observed distinct rings. The most natural source of concentrated dust is molecular clouds. However, only \wsim20-30\% of the ISM in mass is in molecular clouds \citep{Draine2011}, and the neutral \HI\ regions also carry large amounts of dust. We do not possess high-resolution images of the region in 21 cm, and we relied only on low-resolution surveys to estimate the atomic contribution to the total column density. As seen in Fig.~\ref{fig:NHIfit}, for most molecular clouds, there is an associated peak in the \HI\ spectrum, certainly for the most important clouds MC-73/MC-80 and MC-35/MC-39 complex. Since dust should also exist in regions other than the molecular clouds, we added two types of contributions from $N_{HI}$ to overall $N_{H}$, a contribution that is proportional to the molecular column ($N_{H,W}$), as well as a uniform distribution added to all pixels ($N_{H,U}$, see \S~\ref{subsec:imgen}). An azimuthal, or imaging-based analysis as conducted in this work would be less relevant if the dust concentration in atomic hydrogen clouds and dust concentration in molecular clouds were highly uncorrelated. Perhaps, the relatively poor reduced $\chi^{2}$ we obtained in the azimuthal fits can be explained by the presence of additional dust in specific positions due to atomic H clouds (e.g. in the first quadrant North-East in Fig.~\ref{fig:wedfits}, corresponding to wedges 20-22 and 38-40 shown in Fig.~\ref{fig:wedges}).

Another variable that one needs to take into account is the variation in the dust-to-gas ratio, both within the clouds and between the clouds \citep{Giannetti2017}. In our case, neither is allowed to vary. In K18, allowing some variation in the $N_{H}$ of clouds resulted in better radial fits, however, the variations in K18 may also be caused by the shape of the clouds that we try to take into account in this work. Allowing such a variation will certainly reduce the $\chi^{2}$ in the azimuthal fits, but independently varying all normalization with 15 clouds will possibly lead to overfitting, not to mention the prohibitively long time to generate all images to check all variations. 

\subsubsection{Ambiguity in cloud distance estimation and the error in source distance estimation}
\label{subsec:errorest}

In this work, and in K18, we used the kinematic distance estimation method which assumes that clouds move on circular velocities around the Galactic center. The calculation using the relation between the cloud's heliocentric distance and the line-of-sight velocity $v_{lsr}$ results in two solutions, with the so-called near- and far-distance estimates \citep{Schmidt1957, Kolpak2003}. Additional analysis is required to resolve this kinematic distance ambiguity (KDA), such as comparing apparent radii with radii estimated through molecular line widths \citep{Solomon87}, comparing apparent radii with radii estimated using molecular line width and surface density \citep{MivilleD17}, using \HI\ absorption \citep[][and references therein]{Riener2020}, reddening of foreground and background stars \citep{Schlafly2014}, association with spiral arms \citep{Reid2016}, or trigonometric parallaxes \citep{Reid2019}. In this work, rather than trying to resolve the ambiguity with additional methods, we let being in the near- or far-distance estimate as a free parameter and let the azimuthal and radial fitting algorithms pick the best estimate. For our best distance estimate, the results are encouraging, we were able to constrain the placement of 7 out of 15 clouds. Many of the clouds that we could not constrain are weak emitters (see Table~\ref{table:cloudpar} and Figs.~\ref{fig:wedcloudstats}). They also are low $v_{lsr}$ clouds, indicating that they are either behind the source (far-distance estimate), or having such a low $x$ value that (near-distance estimate) at the time of observation, the input flux from the source is low (see Eqn.~\ref{eqn:deltat}). More success could have been achieved in determining the distances of clouds with a series of sensitive X-ray observations during the decay that would allow for picking contributions of low $x$ values with brighter input flux. Finally, if we compare our cloud placements with K18, MC-73, MC-80, MC-100, and MC-105 match, and are consistent with the estimation from the $\sigma-R$ relation \citep{Solomon87}. 

Apart from the KDA, the presence of spiral density waves introduces non-circular motions in the gas, causing deviations from the simple rotation curve base estimation \citep{Ramon-Fox2017}. These radial and azimuthal peculiar velocity components introduce dispersions typically less than 10 km/s, however, they can be as high as 20 km/s in specific regions within 5 kpc of the Galactic center and within the Norma arm \citep{Reid2019}. Therefore a systematic uncertainty of 0.5 kpc exists due to the peculiar velocities of molecular clouds.

The second source of uncertainty is the Galactic rotation curve used in determining kinematic distance estimates. While we used \cite{Reid14} A5 model to be consistent with K18, using other models such as those given in \cite{Reid09, Reid2019, Brand93} resulted in variations of 0.2-0.5 kpc for the far distance estimate of the main cloud, MC-80. 

Finally, there is the uncertainty of the order of 0.1 kpc arising from the radial fitting. The $\chi^{2}$ changes steeply as the source distance is varied with respect to the given cloud distances. Overall, the sum of all systematic and statistical errors is 1 kpc dominated by uncertainties in determining the actual cloud distances.

\subsubsection{Extinction and local absorption}
\label{subsec:extinction}

For sources in the Galactic Plane, it is difficult to estimate the total column density. One approach relies on X-ray spectral fits, but the $N_{H}$ obtained through the fits depends on the model chosen, the properties of the instrument, the abundance, and the cross sections. As already mentioned in \S~\ref{subsec:fu}, the column density is known to vary in this source. Interestingly, \swift\ observations used here (Table~\ref{table:obs}) show significant variations despite using the same instrument, abundances, and cross sections, leaving changes local to the source as the most likely cause for the variability within a given dataset. Such changes could then be attributed to absorption from the variable accretion disc outflow known to be present in 4U~1630$-$47 (see below), or alternatively, to changes in the source continuum during the outburst that are not well captured by the model. 

One can also try to calculate the column density by adding the contributions from the molecular and neutral H clouds as we used in our formulation, however, both contributions will underestimate the $N_{H}$ as the emission will be optically thick. Our study suffers from a discrepancy between the two methods. We used a $diskbb$ model which is relevant for \fu\ in the soft state, and also used $wilm$ abundances \citep{Wilms00} and $vern$ cross-sections \citep{Verner96} in the fits, which resulted in typical $N_{H}$ of \wsim13 $\times 10^{22}$ cm$^{-2}$ (see Table~\ref{table:obs}). 
However, X-ray spectral fitting throughout the literature has resulted in a relatively large range of $N_{H}$ values \citep[$\sim$5-15$\times 10^{22}$ cm$^{-2}$, e.g.][]{DiazTrigo14, Gatuzz19, Neilsen14, Tomsick14}. While the lowest values from the range in the literature are roughly consistent with the maximum value we obtain from molecular clouds and atomic \HI\ (6 $\times 10^{22}$ cm$^{-2}$, see Table~\ref{table:radfits}), the $N_{H}$ value used in this paper is rather at the highest end of the range.

 As a result, our generated images are much brighter than observed as evidenced by the low normalization factors in the case of 11.5 kpc. To put this discrepancy in another way; when we calculate unabsorbed fluxes, we use a much higher $N_{H}$ compared to that used in the absorption factor in Eqn.~\ref{eqn:extinction}. Our aim is not accurately determining the extinction; therefore, all the discrepancies here are adjusted by the normalization factor. However, this choice may have some effect on the energy-dependent azimuthal distribution, as extinction by the intervening and post-scattering clouds is calculated separately. 

For the model used in this paper, there will be less discrepancy if metal abundances are increased. In fact, if we use $angr$ abundances \citep{Anders89}, while the unabsorbed fluxes remain the same, the $N_{H}$ from the spectral fits reduce to \wsim10 $\times 10^{22}$ cm$^{-2}$ and the generated images become dimmer because of the increased absorption cross section, such that the normalizations are now closer to 1. Another interesting possibility is that the different values observed in the fits to X-ray spectra are partially due to variable local absorption at the accretion environment of the source. \fu\ is a high inclination system showing strong, highly photoionised absorption from a disc wind with column densities above \wsim10 $\times 10^{22}$ cm$^{-2}$ \citep[e.g.][]{Kubota07, DiazTrigo14, Gatuzz19}. While so far the presence of a colder component of this wind could not be demonstrated due to the high extinction in the line of sight, the presence of cold winds has been established already in a few X-ray binaries and the co-existence of the hot and cold components is a subject of intense research \citep[][]{MunozDarias19,MunozDarias22}. 

Recent work by \cite{Kushwaha2023} shows that some $NICER$ spectrum of this source could only be fitted if partial absorption ($pcfabs$) is added to the model, with almost equal amounts of $N_{H}$ in $pcfabs$ and $tbabs$. When we tried adding $pcfabs$ to the \swift\ observations discussed in \S~\ref{subsec:xray}, our fits improved with substantial $N_{H}$ in $pcfabs$, but the improvement was not statistically significant. While we cannot confirm the presence of local absorption with the quality of our spectral data, it is an intriguing possibility that would reduce our discrepancy in the normalization factors and it is also supported by other observations.

\begin{figure}
\includegraphics[width=\columnwidth]{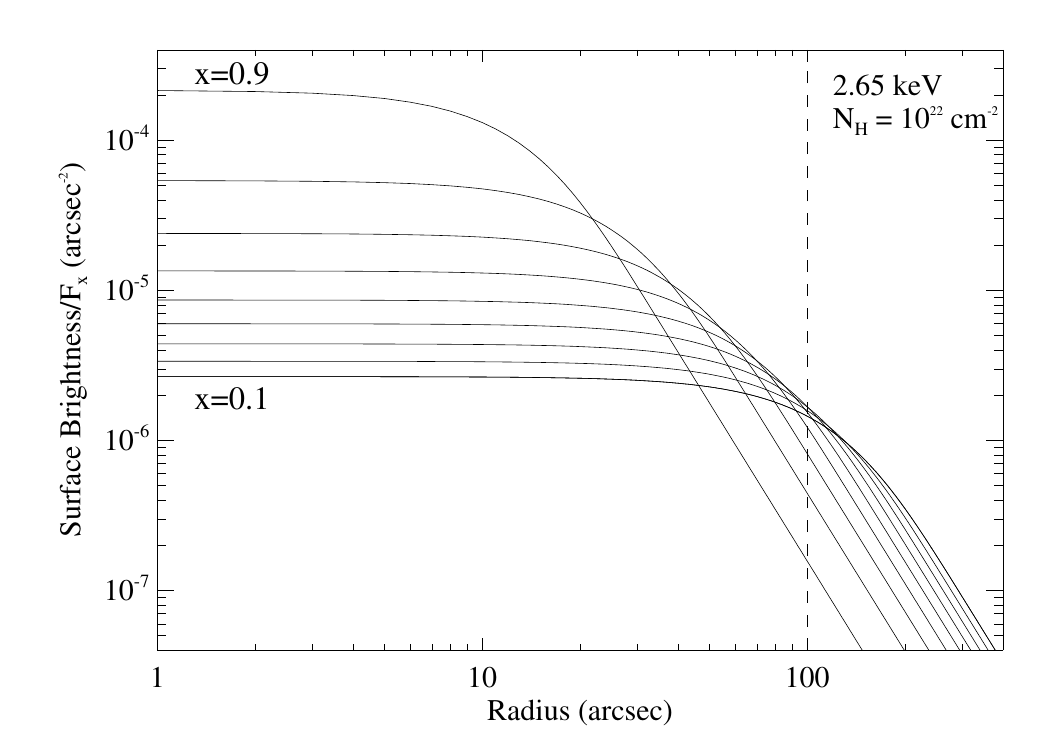}
\caption{\label{fig:normint}
The surface brightness divided by X-ray flux ($norm\_int$) obtained from \textit{Newdust} fits for 10 $x$ values, from 0.1 to 0.9. The values are calculated at an average energy of 2.65 keV (E2) and for an $N_{H}$ of $1 \times 10^{22} \rm cm^{-2}$. The dashed line is at 100\arcsec, the radius at which the ring profile is rising quickly.
}
\vspace{0.5 pt}
\end{figure}

\subsubsection{Scattering cross-section}
\label{subsec:newdust}

In K18, we used a generic power-law to represent the cross section as a function of scattering angle. However, while a power law is a good fit for the scattering cross section for large angles, for smaller angles, the cross-section flattens. This is evident in \textit{Newdust} calculations of the normalized intensity ($norm\_int$) as shown in Fig.~\ref{fig:normint} as the cross-section flattens for $x$ values less than 0.5 close to 100\arcsec, the radius at which the halo brightens significantly. We note that for the main halo, between 100\arcsec\ and 250\arcsec, \textit{Newdust} cross sections as a function of scattering angle can be represented by a power-law, and therefore K18 results are still validated. However, fits extending to lower $\theta$ are affected more significantly.

The other advantage of using \textit{NewDust} instead of a generic power-law is obtaining the exact cross-section, where in K18 an unknown normalization factor was used. We did not consider dust models other than the default in \textit{NewDust}. 

We stress that the calculations in \textit{Newdust} are for an optically thin medium, whereas our column densities indicate that this may not be a valid approximation and there may be a significant number of double scatterings. We utilized the $REFLEX$ ray-tracing code \citep{Ricci2023} to check the ratio of double and single scatterings as a function of column density. We first simulated a geometrically thin (0.2 pc) sheet of gas and dust. We confirmed that the $REFLEX$ generated image's surface brightness profile matches that of \textit{Newdust} for the same energy distribution and $x$ value. We then set up the $REFLEX$ parameters with a monochromatic source at 2.7 keV, a cylindrical cloud of thickness 200 pc at a distance of 650 pc away from the source which is at 11.5 kpc. The camera size and resolution mimicked the \apex\ field of view and resolution. This configuration very roughly represents MC-80 for our best distance case.  We varied the column density between 0.5 -- 10 $\times 10^{22} \, \rm cm^{-2}$ and obtained the ratio of double scatterings to all dust scatterings as 1.4\%, 2.8\%, 11.7\%, and 18.5\% at 0.5, 1., 5., and 10$\times 10^{22} \, \rm cm^{-2}$ respectively. Given that the actual column density is very likely to be larger than 5 $\times 10^{22} \, \rm cm^{-2}$, our approach does not take into account double scatterings that are between 10-15\% of all interactions recorded. The net effect of double scatterings would be loss of flux if the double scattering removes the photon from the field of view, and the smoothing of the halo image if the photon stays, which could be another reason that our fits require a strong uniform dust distribution.  

\section{Summary and future work}

Using high-resolution mm and X-ray band images from \apex, and  \chandra, we developed a method to determine the distance to the peculiar black hole binary source \fu. The method includes a machine learning formalism to determine 3D representations of molecular clouds and uses the X-ray flux history and dust scattering basics to generate synthetic images of the dust scattering halo for all combinations of near and far distance estimates of molecular clouds in the line of sight.

The study confirmed the previous distance estimate of 11.5 kpc for the source, and provided strong constraints on the near and far placement for 7 out of 15 molecular clouds. Other distance estimates, while providing a low $\chi^{2}$ in our fitting algorithm, were ruled out when the full \chandra\ image, $^{12}$CO spectra from the past surveys, and the total extinction were taken into account. 

While molecular cloud images and their radial and azimuthal distributions were taken into account in earlier work \citep[e.g.][]{Heinz15}, this is the first time that very high-resolution mm images are utilized to produce synthetic images for direct comparison with the observed X-ray halo. The results are promising and can be improved further in future studies. High-resolution maps and spectra of neutral hydrogen gas could highly complement the method as the results indicate as much contribution from neutral hydrogen clouds as the molecular clouds. This may also help with obtaining a better grip on the total $N_{H}$ in the line of sight. Our method keeps the overall normalization free because we cannot accurately determine the total hydrogen density, whereas this parameter comes into our equations both in determining dust scattering intensity and absorption. A future study could tackle finding the right combination of H density and absorption cross sections to generate consistent images in more than one energy band. 

This analysis may be extended to other sources with existing halo images. The X-ray imaging resolution does not have to be super-high, as the beam width of \apex\ is 30\arcsec. However, the timing of observations and the sensitivity are important, when the source is bright, the X-ray telescope PSF and the halo image may be hard to disentangle.

With future high-resolution and high-statistics X-ray images \citep[e.g. using $Lynx$ or $AXIS$;][]{Gaskin2019, Reynolds2023}, and known source distance, the problem can be reversed, and one can determine the distances to the molecular clouds with good accuracy to test Galactic rotation curve models, as well as how much the molecular clouds stick to the distances given by their radial velocities.

\section*{Acknowledgements}  
This publication is based on data acquired with the Atacama Pathfinder Experiment (APEX) under programme IDs 0107.F-9323 and 0110.F-9305. APEX is a collaboration between the Max-Planck-Institut fur Radioastronomie, the European Southern Observatory, and the Onsala Space Observatory. Swedish observations on APEX are supported through Swedish Research Council grant No 2017-00648. This publication has made use of MAXI data provided by RIKEN, JAXA and the MAXI team. The authors thank Sebastian Heinz, Lia Corrales, and Stephane Paltani for useful discussions. The authors thank the anonymous reviewer for their valuable suggestions and corrections that have improved the clarity and quality of this publication.

\section*{Data Availability}

The \chandra\ and \swift\ data underlying this article are available in HEASARC \url{https://heasarc.gsfc.nasa.gov/docs/archive.html}, and the \maxi\ data are available at \url{maxi.riken.jp}. The \apex\ data are available at ESO archives \url{https://archive.eso.org/wdb/wdb/eso/apex/form}. The generated images in this article will be shared on reasonable request to the corresponding author.

\bibliographystyle{mnras}
\bibliography{refs}

\appendix

\section{3D cloud representation algorithm}

In this section, we explain the algorithm used to determine the 3D shapes of interstellar molecular clouds along the line of sight to the source object.
The goal is to reconstruct the 3D morphology of molecular clouds using high-resolution 
 
maps from \apex. These data provide detailed insight into the cloud distribution along the line of sight. The main challenges are the overlap of clouds in velocity space and the narrow field of view, which limits the capture of full cloud structures. Standard clump identification algorithms such as ClumpFind \citep{williams1994determining}, GaussClumps \citep{stutzki1990high}, FellWalker \citep{berry2015fellwalker}, Reinhold \citep{berry2007cupid}, and Dendrograms \citep{rosolowsky2008structural} were tested but were unsuitable for our data, as they either overestimated or failed to identify meaningful clumps due to these constraints. Hence, an alternative approach was adopted.

\begin{figure}
\includegraphics[width=\columnwidth]{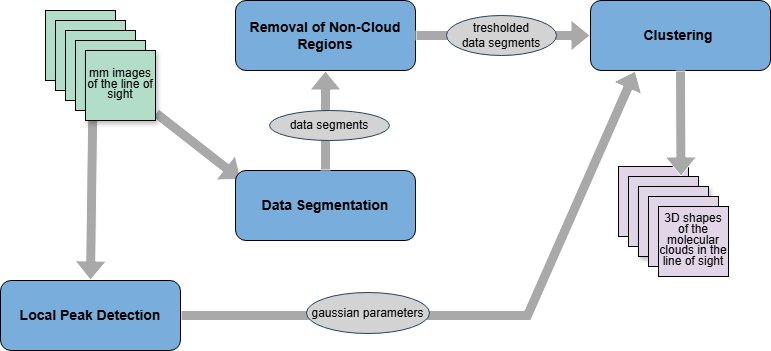}
\caption{\label{fig:method_summary}
Illustration of the data processing pipeline for molecular cloud detection, involving local peak detection, segmentation, removal of non-cloud regions, and clustering.}
\vspace{0.5 pt}
\end{figure}

The approach consists of four core stages: local peak detection through Gaussian decomposition, segmentation of the data, threshold to identify cloud regions, and clustering data points into clouds. (see Fig.~\ref{fig:method_summary}) 

\subsection{Initial Gaussian Fits and Cloud Identification}

In the first step, spectra along the line of sight are analyzed by integrating pixel values across all velocity channels. Integrated spectra is fitted with Gaussian functions, under the assumption that each Gaussian component corresponds to one molecular cloud. A reduced chi-square minimization is employed to determine the minimum number of Gaussians required to fit the data ($\chi^2 \approx 1$) as 15, which is recorded as the number of clouds (see Fig.~\ref{fig:decomposition}).

For each fitted Gaussian, the key parameters—amplitude, central velocity ($\mu$), and velocity dispersion ($\sigma$)—are extracted. The velocity interval for each cloud is defined as the range from $\mu - 3\sigma$ to $\mu + 3\sigma$, capturing approximately 99.73\% of the cloud's intensity. These boundaries are then rounded to the nearest 0.5 or whole number velocity unit to maintain consistency with the data's resolution. This approach results in some overlap between adjacent clouds, which is addressed in later steps. The peak, or center, of each cloud is identified by selecting the velocity value at the Gaussian peak ($\mu$) as the cloud's central velocity. In the spatial dimensions ($x$, $y$), the cloud's center is determined by recording the coordinates of the pixel with the maximum intensity at this velocity value. 

\subsection{Data Segmentation}

In the second step of the algorithm, the data is divided into five distinct segments using the velocity intervals identified in the previous step. Clouds with overlapping velocity intervals are grouped. For most cloud groups, no overlapping velocities were present. However, in one case where a minor overlap was observed, the overlapping region was divided evenly between the two groups.
This segmentation not only reduces the data size, enhancing the efficiency of both the thresholding and clustering processes but also allows for more precise application of thresholds in each segment. By minimizing overlap and grouping similar velocity regions, the segmentation ensures that subsequent clustering steps are performed more effectively.

\subsection{Thresholding}

\begin{figure}
\includegraphics[width=\columnwidth]{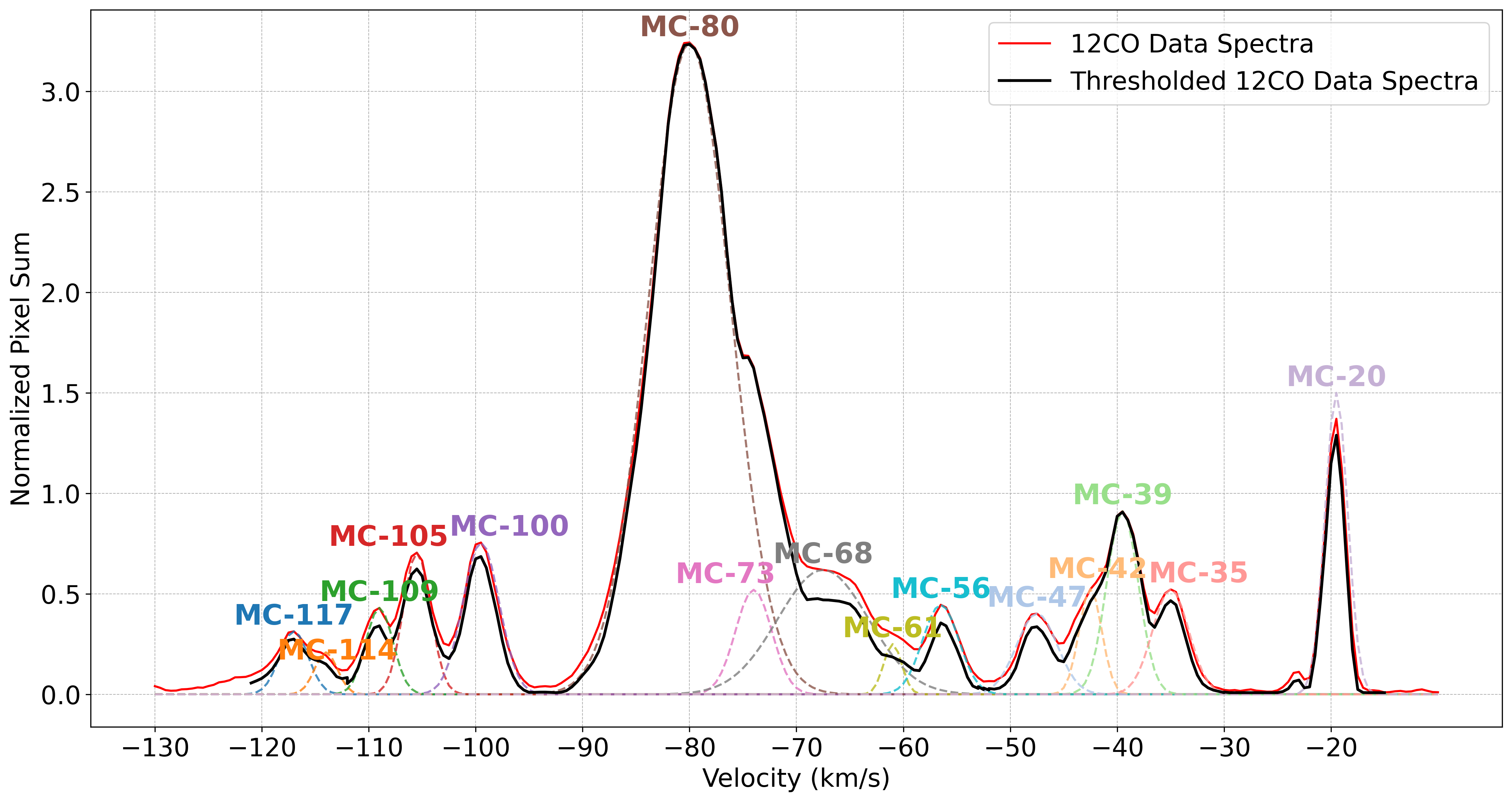}
\caption{\label{fig:decomposition}
Velocity spectrum created by summing pixel intensities across velocity channels (-130 km/s to -10 km/s). The red line shows the normalized original data, the blue line represents Otsu-thresholded data, and the dashed lines are Gaussian fits representing 15 molecular clouds along the line of sight.}
\vspace{0.5 pt}
\end{figure}

In this step, Otsu’s thresholding method is applied to each data segment to separate cloud-occupied and non-cloud regions based on pixel intensity. Otsu’s algorithm automatically determines an optimal threshold by minimizing within-class variance  \citep{otsu1975threshold}. This method is widely used in astronomical image segmentation, including radio astronomy for thresholding \citep{masias2012review} \citep{xu2024surveying}. Each segment is thresholded separately, allowing for region-specific thresholds that improve accuracy. The non-cloud regions, which fall below the computed threshold, are discarded, and the cloud regions are passed to the clustering step (see Fig. ~\ref{fig:decomposition}).

\subsection{Clustering}

In the final step, clustering is performed on the thresholded data segments, where each cluster corresponds to a molecular cloud. A modified version of the Mean Shift algorithm is applied after normalizing the data. The number of clusters and their centers, determined in earlier steps, are used as inputs for the clustering. Hyperparameters are fine-tuned using error functions based on the cloud properties. After obtaining the optimal clustering parameters, the results are post-processed to generate the 3D shapes of the clouds.

\paragraph{Data Preprocessing}

Each pixel in the thresholded segments is treated as a data point with attributes (x, y, velocity, intensity), and all features are normalized to the [0, 1] range using Min-Max scaling to prepare for the Mean Shift clustering.

\paragraph{Clustering Algorithm}

Each data segment is clustered with Mean-Shift algorithm into a predefined number of clouds, as determined during Gaussian decomposition. The Mean-Shift algorithm, which normally starts with random cluster centers, is applied here using the predetermined peaks as initial cluster centers. The algorithm iteratively shifts these centers toward regions of higher data density, converging at local maxima, where clusters form around the densest regions \citep{cheng1995mean} \citep{fukunaga1975estimation}.  Mean-Shift has been widely used in image segmentation tasks and also in remote sensing \citep{comaniciu1999mean} \citep{zhou2011mean} \citep{mohan2013importance}.
While Mean-Shift typically performs hard clustering, assigning each data point to a single cluster, we adapted it to soft clustering using a Gaussian kernel to handle overlapping clouds. This allows data points to belong to multiple clusters with associated probabilities. To refine these probabilities, the alpha and beta parameters are introduced. The alpha parameter sets a lower threshold, redistributing small probabilities into significant cluster memberships, while the beta parameter caps higher probabilities, treating values above the threshold as fully assigned to a cluster. These modifications help manage overlapping regions and prevent the formation of insignificant clusters.

\paragraph{Hyperparameter Tuning and Error Functions}

Several hyperparameters require tuning in the clustering algorithm, including bandwidths for each cluster, feature scales, and the alpha and beta parameters. Hyperparameter optimization is performed using a combination of the Silhouette index \citep{rousseeuw1987silhouettes} and two custom error functions.
The tuning process begins with gradient descent, starting from default initial values, and refines the parameters based on the Silhouette index. Final fine-tuning is conducted using the two error functions, which are explained below. The final parameters are manually selected from the best results for each data segment.

\subsection{Error Functions}

Two error functions were applied in the hyperparameter tuning process to evaluate the spatial properties of the cloud shapes. These functions assess shape consistency and smoothness to optimize clustering results.

\paragraph{Shape Smoothness Error}

This error function evaluates the smoothness of cloud shapes by comparing consecutive velocity segments. The difference between each segment and its neighboring segments is computed, and the sum of these differences is normalized by the number of velocity segments per cloud. The total error is then averaged across all clouds. The formulation of the error function is given below:

\begin{equation}
\text{error} = \frac{1}{N} \sum_{c} \Bigg[ \frac{1}{S_c} 
\sum_{v} \big(2 \cdot I_v - I_{v+0.5} - I_{v-0.5}\big) \Bigg]
\end{equation}

Where $N$ is the total number of clouds, $S_c$ is the size of cloud $c$, and $I_v$ is the cloud segment at velocity $v$.

\paragraph{Eccentricity Consistency Error}

This error function measures the consistency of cloud shapes by comparing the 
eccentricity of each velocity segment to the central segment. The eccentricity is determined using the eigen-ratio of the elliptic shapes. The squared difference between the central and other segments is summed, normalized by the number of segments, and averaged across all clouds.
 The formulation of the error function is given below:

\begin{equation}
\text{error} = \frac{1}{N} \sum_{c} \Bigg[ \frac{1}{S_c} \sum_{v} \big(R_c - R_v\big)^2 \Bigg]
\end{equation}

Where $N$ is the total number of clouds, $S_c$ is the size of cloud $c$, $R_c$ is the center eigen-ratio of cloud $c$, and $R_v$ is the eigen-ratio at velocity $v$.

\subsection{Data Postprocessing}

For each pixel with a soft cluster assignment, the pixel is divided into separate data points for each cluster. Each pixel is split into new data points, with their intensities scaled by the probability of belonging to each cluster. Once all pixels are processed, the final cloud structures are formed (see Fig. ~\ref{fig:3drep}). Finally, a Gaussian smoothing function is applied along the velocity axis of each pixel of each cloud. This redistributes the pixel brightnesses around the centroid slice of each cloud while maintaining the total summed intensity for each $(x_p,y_p)$, ensuring the cloud remains within a specific thickness (in pc) whilst maintaining the overall intensity. The thicknesses of clouds are determined from the near and far distance size estimates given in \cite{MivilleD17}, similar to what is done in K18.

\end{document}